\documentclass[acmsmall]{acmart}

\setcopyright{cc}
\setcctype{by-nc-nd}
\acmJournal{PACMMOD}
\acmYear{2026} \acmVolume{4} \acmNumber{6 (SIGMOD)} \acmArticle{379}
\acmMonth{12} \acmDOI{10.1145/3856374}

\usepackage{multirow}
\usepackage{algorithm}
\usepackage{makecell}
\usepackage{algpseudocode}
\usepackage{color}
\usepackage{tcolorbox}
\usepackage{lipsum}
\usepackage{tcolorbox}
\usepackage{booktabs}
\tcbuselibrary{breakable}
\tcbuselibrary{skins}

\AtBeginDocument{%
  }

\setcopyright{acmlicensed}
\copyrightyear{2026}

\newcommand{\modelname}{TEAR}

\newcommand{\revision}[1]{{#1}}

\newcommand{\eat}[1]{}

\begin{document}
\received{18 April 2026}
\received[revised]{20 August 2026}
\received[accepted]{13 September 2026}

\title{TEAR: Table Extraction with Attribute Recommendation from Texts via Large Language Models}

\author{Tong Li}
\orcid{0009-0000-6486-8528}
\affiliation{
  \institution{Hong Kong University of Science and Technology}
  \city{Hong Kong SAR}
  \country{China}
}
\email{tlice@connect.ust.hk}

\author{Shuye Ding}
\orcid{0009-0004-7105-3377}
\affiliation{
   \institution{Hong Kong University of Science and Technology}
  \city{Hong Kong SAR}
  \country{China}
}
\email{sdingah@connect.ust.hk}

\author{Jiachuan Wang}
\orcid{0000-0001-6473-8221}
\authornote{Corresponding author.}
\affiliation{
  \institution{University of Tsukuba}
  \city{Tsukuba}
  \country{Japan}
}
\email{wangjc@slis.tsukuba.ac.jp}

\author{Yongqi Zhang}
\orcid{0000-0003-2085-7418}
\affiliation{
  \institution{Hong Kong University of Science and Technology (Guangzhou)}
  \city{Guangzhou}
  \country{China}
}
\email{yzhangee@connect.ust.hk}

\author{Shuangyin Li}
\orcid{0000-0001-6404-3438}
\affiliation{
  \institution{South China Normal University}
  \city{Guangzhou}
  \country{China}
}
\email{shuangyinli@scnu.edu.cn}

\author{Lei Chen}
\orcid{0000-0002-8257-5806}
\affiliation{
  \institution{Hong Kong University of Science and Technology}
  \city{Hong Kong SAR}
  \country{China}
}
\email{leichen@cse.ust.hk}

\author{Bo Li}
\orcid{0000-0003-2083-9105}
\affiliation{
  \institution{Hong Kong University of Science and Technology}
  \city{Hong Kong SAR}
  \country{China}
}
\email{bli@cse.ust.hk}


\begin{abstract}
Table extraction from texts is an important task for information systems, and recent approaches that prompt large language models (LLMs) with instructions have drawn great attention for their strong performance. Existing works have assumed the input texts to be table descriptions or specialized documents. However, these efforts have largely overlooked another prevalent category of texts, commonly found in news reports and social media: naturally occurring texts. Extracting tabular information from such texts poses two distinct challenges. First, high variability and the absence of explicit structural cues make fixed heuristic LLM prompts limited in precisely delineating extraction boundaries. Second, manually predefined schemas cannot capture open-ended, unseen attributes in naturally occurring text. In this paper, we propose a framework, TEAR, to address these challenges. It comprises two synergistic workflows: a Table Extraction Workflow that dynamically adapts instructions to overcome the limitation of heuristic instructions, and an Attribute Recommendation Workflow that discovers new attributes from texts to complement the heuristic schema. To our knowledge, TEAR is the first framework that supports automated text-driven attribute recommendation, enabling exploratory schema design for table extraction. To evaluate TEAR, we establish the benchmark for table extraction and attribute recommendation on naturally occurring texts, including two real‑world datasets, manual annotations, appropriate metrics, and baseline comparisons. Experiments show that TEAR achieves state‑of‑the‑art performance on both tasks, and the recommended attributes effectively enhance extraction performance in exploratory scenarios.
\end{abstract}

\ccsdesc[500]{Information systems~Information extraction}
\ccsdesc[500]{Computing methodologies~Information extraction}

\keywords{table extraction, attribute recommendation, text-to-table, large language models}


 \maketitle

\section{Introduction}
\begin{figure*}
    \centering
    \includegraphics[trim=0 160 0 0, clip, width=\linewidth]{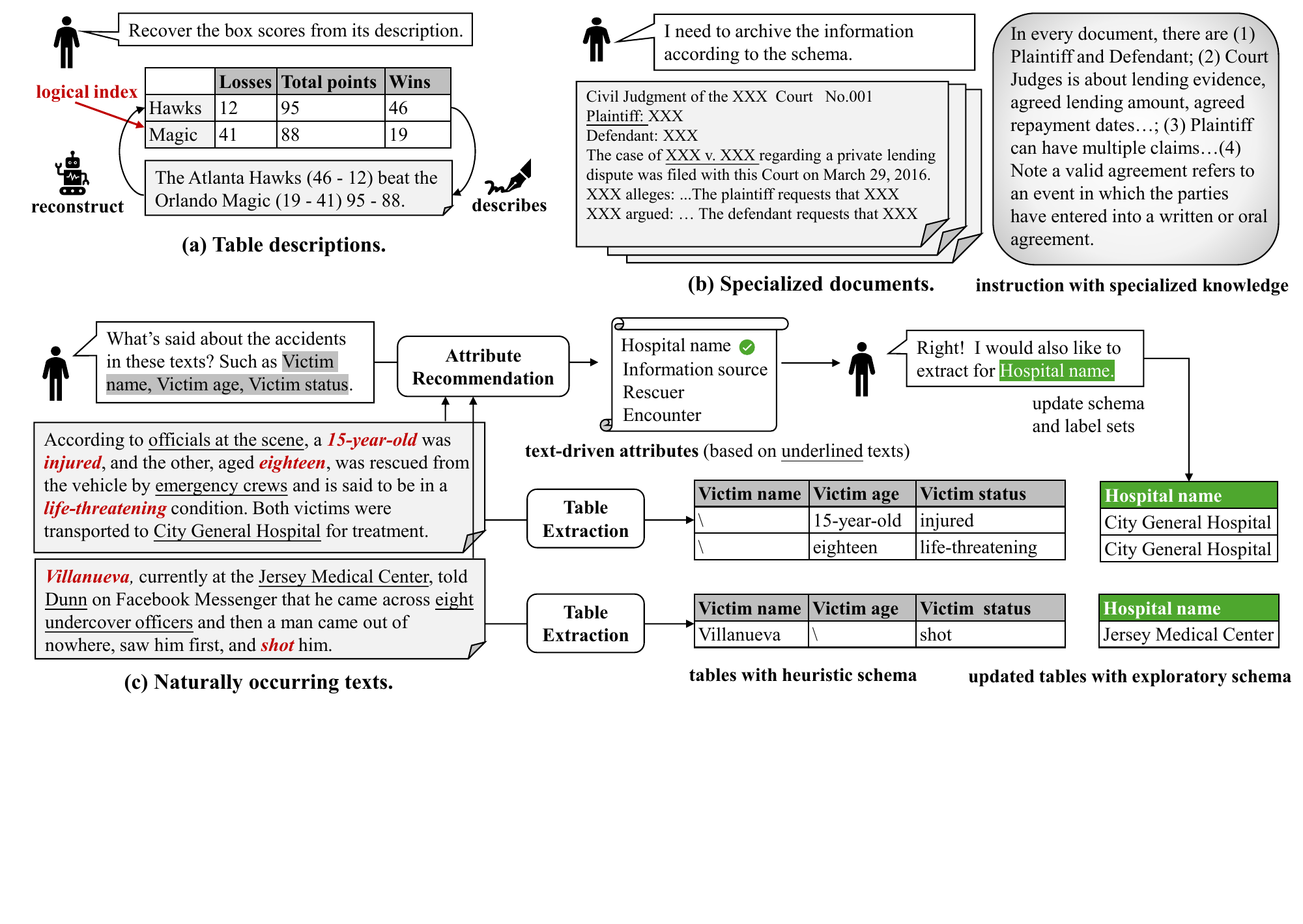}
    \caption{Table extraction from different categories of input texts. }
    \label{fig:source-text}
\end{figure*}
Table extraction from texts, also known as \textit{text-to-table}, is an emerging task focused on identifying semantic values from unstructured texts and organize them into tabular format~\cite{wu2022text}, which unlocks the practical utility of texts for a wide range of downstream applications \cite{jiang2024tkgt, zhao2024medt2t, newman2024arxivdigestables, li2025adapting}, as well as simplifies data management~\cite{fagin2016relational, zhou2022survey, zhang2025survey, yuan2023survey}. 

Previous works have developed into two lines, focusing on different categories of input texts. 
The first category is \textit{table descriptions}, which are manually written or artificially generated according to well-defined tables, such as the NBA game summary and their original box scores in Figure~\ref{fig:source-text}(a). 
In this line of research, the tables exist natively, and description text can be obtained in bulk~\cite{lin2023survey, novikova2017e2e, wiseman2017challenges, bao2018table}.  
Thus, with paired (text, table)s, researchers adopt a supervised paradigm, training generation models end-to-end to reconstruct the original table from its description~\cite{wu2022text, li2023sequence, pietruszka2024stable}.  
However, their input texts are generated under control and are dominated by the pre‑existing tables, which are limited in reflecting the true difficulty of extracting from real‑world texts.
%

The second category is \textit{specialized documents}, such as legal documents~\cite{jiang2024tkgt, chai2025doctopus} and biographies~\cite{lebret2016neural, chai2025doctopus}.
These documents do not come with pre‑aligned tables and involve more complicated contents, making annotating sufficient paired data for supervised learning expensive.
Alternatively, researchers have shifted toward agentic extraction~\cite{jiang2025tst, jiang2024tkgt, chen2024decomposed} powered by large language models (LLMs).
This paradigm is known as in‑context learning (ICL)~\cite{dong2024survey}, where the user depicts the task as instruction prompts that guide the LLMs to locate and extract relevant information without task-specific fine-tuning.
The effectiveness of ICL stems not only from LLMs' general semantic understanding, but also from how these documents are composed to facilitate information retrieval.
Specifically, specialized documents usually follow established writing conventions to convey predefined knowledge, offering structural cues for human readability, which the LLM can also recognize and exploit. 
For example, in Figure~\ref{fig:source-text}(b), the prefix ``Plaintiff:'' or pattern ``Party A v. Party B'' are reliable cues that help readers and LLMs quickly identify important information. 

\paragraph{\underline{Our focus}} 
Although effective for their assumed input texts, previous works have overlooked the extraction demands for \textit{naturally occurring texts}~\cite{liddy2001natural, lester2017naturally}, the unscripted language produced for daily human communication, pervasive in news reports, social media discourse, and customer service interactions.
Ubiquitous naturally occurring texts (n-texts) encapsulate the authentic, unfiltered information that flows through human interaction, a quality inherently absent from curated descriptions or specialized documents. 
Extracting tables from them, therefore, represents a pivotal advance of the field into realistic scenarios.
Unlike previously studied input texts, n-texts are not organized around pre-existing tables or predefined knowledge. 
Instead, they unfold in a fluid, narrative style, exhibiting less literal consistency, which introduces new challenges for extraction.

\textbf{Challenge 1. Ineffective heuristic instruction.}
Resorting to LLM-based in-context learning for n-texts is appealing, given its semantic capability and data efficiency. Nevertheless, n-texts lack stable structural templates or cues in contrast to specialized documents. This high variability means that even carefully crafted instructions are often insufficient to delineate all possible extraction boundaries and convey nuanced task requirements~\cite{qi2024adelie, peng2023does}.

\begin{example}
\label{eg:c1}
Consider the attribute ``Victim status'' in Figure~\ref{fig:source-text}(c). 
Unlike the explicit plaintiff name in Figure~\ref{fig:source-text}(b), its scope is ambiguous: it may encompass only direct casualties (``injured''), or also extend to relocation (``transported''). Such ambiguity is not resolvable via LLM common sense but requires task specification. Exhaustive enumeration of analogous ambiguities in static instruction is not only impractical, but may instead dilute model attention to cause lost-in-the-middle~\cite{liu2024lost}.
\end{example}

\textbf{Challenge 2. Inadequate heuristic schema.}
A crucial step in extraction is understanding what information the texts contain, so as to determine the target schema, i.e., the attributes of interest. 
Existing methods typically work with heuristic schemas provided by human experts~\cite{jiang2024tkgt, jiang2025tst}, which may be satisfactory for their assumed inputs, where experts easily anticipate text contents.
However, n-texts do not adhere to a stable informational template.
Even if one were to navigate massive volumes of such texts and painstakingly summarize the observed contents into attributes, missing some attributes remains a risk.
In other words, users confronting n-texts face an \textit{exploratory scenario}: initially, they can only list some obvious attributes that immediately come to mind, yet still seek to uncover all the relevant attributes that will emerge from the texts ahead.

\begin{example}
\label{eg:c2}
Consider the two texts in Figure~\ref{fig:source-text}(c). Though both are news reports about accidents, one details a rescue effort (``emergency crews''), and the other details a sudden, hostile encounter (``eight undercover officers'').
Such contents diverge more sharply compared to those of the judgment in Figure~\ref{fig:source-text}(b), where each document follows a stable informational template to mention the plaintiff, defendant, case number, etc.
\end{example}

\paragraph{\underline{Our proposals}} 
In this paper, we propose a framework \modelname, short for \textbf{T}able \textbf{E}xtraction with \textbf{A}ttribute \textbf{R}ecommendation, to address the above challenges for naturally occurring texts.

To tackle the dilemma of heuristic instructions of \textbf{Challenge 1}, our idea is to dynamically select and insert \textit{demonstrative examples} into the prompt for each text.
These examples complement the abstract instructions with customized task specifications, enabling the LLM to perform extraction with clearer objectives.
Also, retaining a small set of labeled data to supply such examples offers a practical trade‑off between LLM usability and data efficiency~\cite{qi2024adelie}.
Determining what constitutes an effective demonstration is non‑trivial, because general text similarity~\cite{pradhan2015review} does not fully reflect the task utility of an example, such as its ability to clarify the extraction boundaries of a particular attribute.  
To this end, \modelname~incorporates a proactive module that first performs a fuzzy prediction of what contents are likely to appear in the target table cells, and then retrieves examples that are informative for the specific extraction.

To overcome the limitations of heuristic schemas discussed in \textbf{Challenge 2} and support exploratory scenarios, \modelname~introduces a novel \textit{attribute recommendation} task, which automatically discovers new attributes in texts to assist in refining the prior schema, as in Figure~\ref{fig:source-text}(c).
Although the capabilities of recent LLMs in processing open-ended information~\cite{ahuja2025map,zhang2024extract} render this task possible, naively invoking LLMs and blindly accepting their outputs leaves the system vulnerable.
Intuitively, an effective LLM-based attribute recommendation method should possess further abilities to (i) guide the LLM to propose attributes grounded in texts rather than ad-hoc fabrication; (ii) integrate the LLM's differently articulated discoveries across multiple texts into a global candidate set; and (iii) prioritize the most salient candidates, shielding users from an undifferentiated, exhaustive list. 
Accordingly, we equip \modelname~with components that operationalize these three intuitions.

Finally, we establish an evaluation methodology tailored to this new setting.
Prior work on table descriptions and specialized documents assumes the presence of logical indices to align and compare records~\cite{wu2022text}, which are not applicable to n-texts because they offer no such convenience.
We therefore introduce a new metric for extraction that assesses table semantics without relying on indices, along with a dedicated evaluation protocol for the novel attribute recommendation task.
We accompany this evaluation with two n‑text datasets, comprising 3,375 manually annotated ground‑truth tables spanning two domains.
Extensive experiments on these datasets demonstrate that \modelname~consistently outperforms state‑of‑the‑art baselines on both extraction and recommendation tasks.

To sum up, our main contributions are as follows:

\begin{enumerate}
\item We introduce an LLM-based framework \modelname~ for table extraction with attribute recommendation from naturally occurring texts. 
To the best of our knowledge, it is the first framework that could recommend new relevant attributes derived from texts under exploratory scenarios. 

\item For table extraction, we propose a \textbf{Proactive Demonstration Module} that dynamically retrieves examples for adaptive instructions for each text. 

\item For attribute recommendation, we propose a \textbf{Discovery Mechanism} that frames open‑ended attribute discovery as table expansion to engage the LLM in proposing faithful new attributes, a \textbf{Hybrid Integration Strategy} that resolves duplicate LLM discoveries via diversity checking and contextualized analysis, and a \textbf{Schema Coherence Score} that ranks candidates by their holistic coherence with the initial schema.

\item We establish a new and fair evaluation methodology given the emergent evaluation difficulty from table extraction and attribute recommendation from naturally occurring texts.
\end{enumerate}

In the rest of this paper, we begin with a discussion of related work (Section~\ref{sec:rw}) and the formal problem definition (Section ~\ref{sec:problem_definition}).
We then present \modelname, including an overview (Section~\ref{sec:overview}), the table extraction workflow (Section~\ref{sec:tew}), and the attribute recommendation workflow (Section~\ref{sec:arw}). 
Next, we elaborate on the new evaluation methodology (Section~\ref{sec:eval}). 
Lastly, we report the experiment results (Section~\ref{sec:exp}) and conclusion (Section~\ref{sec:conclusion}).

\section{Related Works}
\label{sec:rw}
\subsection{Table Extraction from Texts}
Existing methods can be classified according to whether model parameters are updated.  
\textbf{1. Supervised paradigm.}
These methods treat table extraction as an end‑to‑end sequence generation task. 
They feed the text sequence into a language generation model~\cite{t5,lewis2020bart} and design different decoding strategies to generate rectangular tables as sequences, 
including row‑by‑row generation~\cite{wu2022text}, parallel row generation~\cite{li2023sequence}, learnable cell ordering~\cite{pietruszka2024stable}, and pointer‑based decoding for the medical domain~\cite{zhao2024medt2t}.
Researchers collect large amounts of paired (text, table) data and update the model parameters to minimize a loss function.
Because these methods rely heavily on the training data distribution, they tend to perform well when the input texts are table descriptions that follow controlled patterns, but they struggle with n-texts, which exhibit higher variability and lack large‑scale training data.
\textbf{2. In-context learning paradigm.}
Recent works employ LLMs as general‑purpose extractors via ICL~\cite{coyne2024large, peng2023instruction, longpre2023flan}.
Researchers convey the task to the LLM of the task through instructions rather than training it, thereby steering model behavior without any parameter updates.
The research focus is on mimicking human‑like behavior by decomposing the extraction task into subtasks (e.g., identifying entities, planning layout, and filling cells), and supplying each subtask with a dedicated instruction~\cite{ahuja2025map, jiang2025tst, jiang2024tkgt}.
These heuristic instructions are coupled with specialized documents and tend to be effective because human experts are themselves proficient at extracting knowledge from such documents.
For instance, TKGT~\cite{jiang2024tkgt} translates human expertise in the legal domain into a knowledge graph and uses it to instruct the LLM. However, it is difficult to guarantee the success of this mode for n‑texts, as the necessary template consistency and domain‑specific heuristics are largely absent.

\subsection{Open Information Extraction}
\label{sec:rw-oie}
\eat{Our attribute recommendation task is related to Open Information Extraction (OIE)~\cite{zhou2022survey} in that both aim to extract undefined information from texts. However, existing OIE methods are not applicable to our task for two reasons.
First, OIE extracts predicates, entities, or n‑ary relations as factual, surface‑level information without summarizing or naming them as attributes~\cite{chen2024sac, zhang2024extract, luo2024text2nkg, liu2021role, guan2019link}. 
For example, OIE may output triples like (“Picasso”, “painted”, “Guernica”) but does not abstract this into a table column “Artwork”.
Second, OIE does not aim to extend a given initial schema, nor does it perform attribute‑level integration or ranking. 
As a result, OIE cannot produce a prioritized recommendation list tailored to a user’s extraction goal as our method does.}
Our attribute recommendation task is related to Open Information Extraction (OIE)\revision{~\cite{zhou2022survey, pai2024survey}.} 
\revision{Among all subareas, the most relevant is novel slot detection~\cite{wu2021novel, wu2022semi, liang2023novel}, since a slot also manifests as a key-value pair. 
However, these works detect the \textit{existence} of a new type, rather than revealing its semantic identity via canonical naming. Further, the slot types they defined are often distinguishable via surface value (e.g., a date vs. a name), hence are insufficient to represent distinct attributes that share similar values (e.g., ``Victim name'' or ``Suspect name'').}
\revision{Other OIE tasks diverge farther from AR.
For example, schema induction and event schema learning~\cite{chen2024sac, zhang2024extract, luo2024text2nkg, liu2021role, guan2019link} discover unseen n-ary tuples representing predicates, entities, and relations, instead of extending known tuples with new dimensions; ontology learning~\cite{wong2012ontology, babaei2023llms4ol} establishes terminology, taxonomies, and axioms that are formal and universal rather than specific to input texts.}
\eat{Crucially, none of these tasks performs attribute-level integration and ranking to produce a recommendation list that extends a given initial schema.}

\subsection{Other Table Extraction Tasks}
\label{sec:rw-other}
Our task aims to identify semantic values in unstructured texts and align them into tables. We acknowledge that several other tasks are also termed “table extraction”, yet their intended application scenarios differ fundamentally from ours.
\textbf{1. Table extraction by format mining.} These methods rely on mining formatting patterns rather than deep semantic understanding, and thus cannot extract attribute values from completely unstructured, free‑form texts. 
For example, web record extraction~\cite{shen2007u, chu2015tegra, zhu2006simultaneous, chen2022web} processes list or detail pages; tables repairing processes CSV files~\cite{christodoulakis2020pytheas, hameed2025repairing} or PDF texts~\cite{singh2024tabularis} with explicit delimiters;
\revision{another recent work~\cite{arora2023language} considers text in heterogeneous data lakes where attributes appear in the fixed form of "\textit{<name>: <value>}".}
\textbf{2. Table extraction by information integration.}
These works focus on reasoning or calculating for the attributes that are not directly stated in the texts. For instance, some count event occurrences~\cite{deng2024text}; others derive attributes such as lifespan or zodiac sign from extracted dates~\cite{chai2025doctopus, dong2024opente}.
Crucially, these approaches do not address the difficulty of extracting atomic attribute values from raw texts, especially highly variable n-texts.
%
\textbf{3. On‑demand table extraction.} 
They extract tables in response to individual user queries rather than an overall schema, which maintains a corpus to locate answers~\cite{chen2024decomposed} or retrieve relevant passages~\cite{chen2024unstructured}. 
\revision{One advantage of our approach is that the pre‑extracted tables enable a broad range of exact SQL‑style operations, not only ad-hoc queries.}


\section{Problem Definition}

\label{sec:problem_definition}

We introduce the related notations.
\revision{The naturally occurring text $W$ is a vanilla word sequence. A \textbf{span}~\cite{fagin2015document} is a subsequence of its words, with $\mathcal{P}(W)$ denoting the \textbf{span space}. 
}
\revision{\begin{definition}[\textbf{Value in Text}]
From the input text,  a valid value for the extracted table is a list of non-overlapping spans. The value space of the given $W$, $\mathcal{A}(W)$, is 
\begin{equation}
\{\langle a_1, a_2, \dots \rangle\mid\forall i, a_i\in\mathcal{P}(W); \forall i<j, a_i\text{ precedes }a_j\}.
\end{equation}
\end{definition}}
Here, a list of spans is used for a value, instead of a single span, because the mentions can be non-consecutive in the naturally occurring text. Our basic settings keep the raw strings as values, allowing the LLM to tokenize and recognize them. For additional specialized usage, one could further convert raw strings into normalized formats by post-processing~\cite{singh2025datavinci}. For example, converting "eighteen" to the numerical "$18$" for actual storing. 
\begin{definition}[\textbf{Schema}]
The extraction schema for naturally occurring text is a set of attributes.
\begin{equation}
    S=\{ s_1, s_2,..., s_{|S|}\}
\end{equation}
\end{definition}
For each attribute in the schema, its name is a string of regular naturalness ~\cite{luoma2025snails}, which contains complete words or acronyms in common usage that are easy to understand by ordinary people and general LLMs. Meaningless or obscure symbols like "Column\_1" or "REV" are not qualified. 

\begin{definition}[\textbf{Table}]
A table $D$ of $n$ records can be viewed as a list, $D=\left[H, R_1,..., R_n\right]$, where $H=[h_1, h_2, ..., h_{m}]$ is the \textit{header region} that consists of $m$ unique attribute names.
$\mathcal{R}=[R_1,...,R_n]$ is the \textit{non-header region}, where the $i$-th record $R_i=[A_{i1},A_{i2},...,A_{im}]$ is a list of $m$ values corresponding to $m$ attributes. 
\end{definition}
Following previous works~\cite{wu2022text, deng2024text}, we define each table as a rectangular data structure with rows as records and columns as attributes. 
If the table \textit{follows} the schema $S$, its header names $H\subseteq S$.
If the table is extracted from text $W$, each value $A_{ij}\in \mathcal{A}(W)$, $i=[1..n]$, $j=[1..m]$. 
We allow some values in the table to be an empty span list, i.e. $|A_{ij}|=0$, indicating not mentioned.  
In a valid table $D$, we assume there is no entirely empty row or column to eliminate dummy structures.
\eat{\begin{equation}    
\begin{aligned}
  &\forall j, \ h_j\neq \textsf{null} \land \sum _{i=1}^{n}|A_{ij}|>0 ,  & \text{(no empty column)}\\
  &\forall i,\  \sum _{j=1}^{m}|A_{ij}|>0. & \text{(no empty row)}\\
\end{aligned}
\end{equation}}


\eat{The dataset may involve several related entity types, and each type corresponds to one output table, such as the "Accident", "Suspect", and "Victim" tables in the incident report dataset~\cite{van2020cacapo}. }

The table extraction (TE) task aims to output stable, task-specific predictions, not only subjectively reasonable ones. The attribute recommendation (AR) task compares the attributes to a referential attribute set. The tasks are defined as follows. 
\begin{definition}[\textbf{Labeled Sample}]
Given a schema $S$, a labeled sample is a pair of text and table $(W^l, D^l)$, where $D^l$ is extracted from $W^l$ and follows $S$. 
\end{definition}
\begin{definition}[\textbf{Table Extraction}] 
Given a schema $S$, labeled samples $\{(W^l_1, D_1^l),(W^l_2, D_2^l),$...$\}$, and unlabeled texts $\{W^u_1, W^u_2,$...$\}$, the method outputs the tables $\{D^{u}_1,D^{u}_2,$...$\}$.
Let $\{\hat D^{u}_1,\hat D^{u}_2,...\}$ be the ground truth, where $\hat D^u_i$ is extracted from $W^u_i$ and follows $S$. 
Given similarity metric $g^{te}$ of any two tables, Table Extraction aim to maximize $g^{te}(D^u_i, \hat D^u_i)$, $\forall i$.
\end{definition}
\begin{definition}[\textbf{Attribute Recommendation}] 
Given an inadequate schema $S$, labeled samples $\{(W^l_1, D_1^l),(W^l_2, D_2^l),...\}$, the unlabeled texts $\{W^u_1, W^u_2,...\}$, and a recommendation budget $k$, the method outputs new attributes $S'$ that $|S'|=k$.
Let $\hat S'$ be the ground truth relevant set of attributes.
Given a recall-based metric $g^{ar}$, Attribute Recommendation aim to maximize $g^{ar}(S', \hat S')$.
\end{definition}

Note that these two tasks share the same input format, except for the recommendation budget $k$. 
Given the input, \modelname accomplishes the TE task like other table extraction systems. 
Under exploratory scenarios with specified $k$, our framework additionally fulfills the AR task in response to the user requirements.

\section{The TEAR framework}
\label{sec:tear}
\subsection{Overview}
\label{sec:overview}
\begin{figure*}
    \centering
    \includegraphics[trim=0 110 30 0, clip, width=\linewidth]{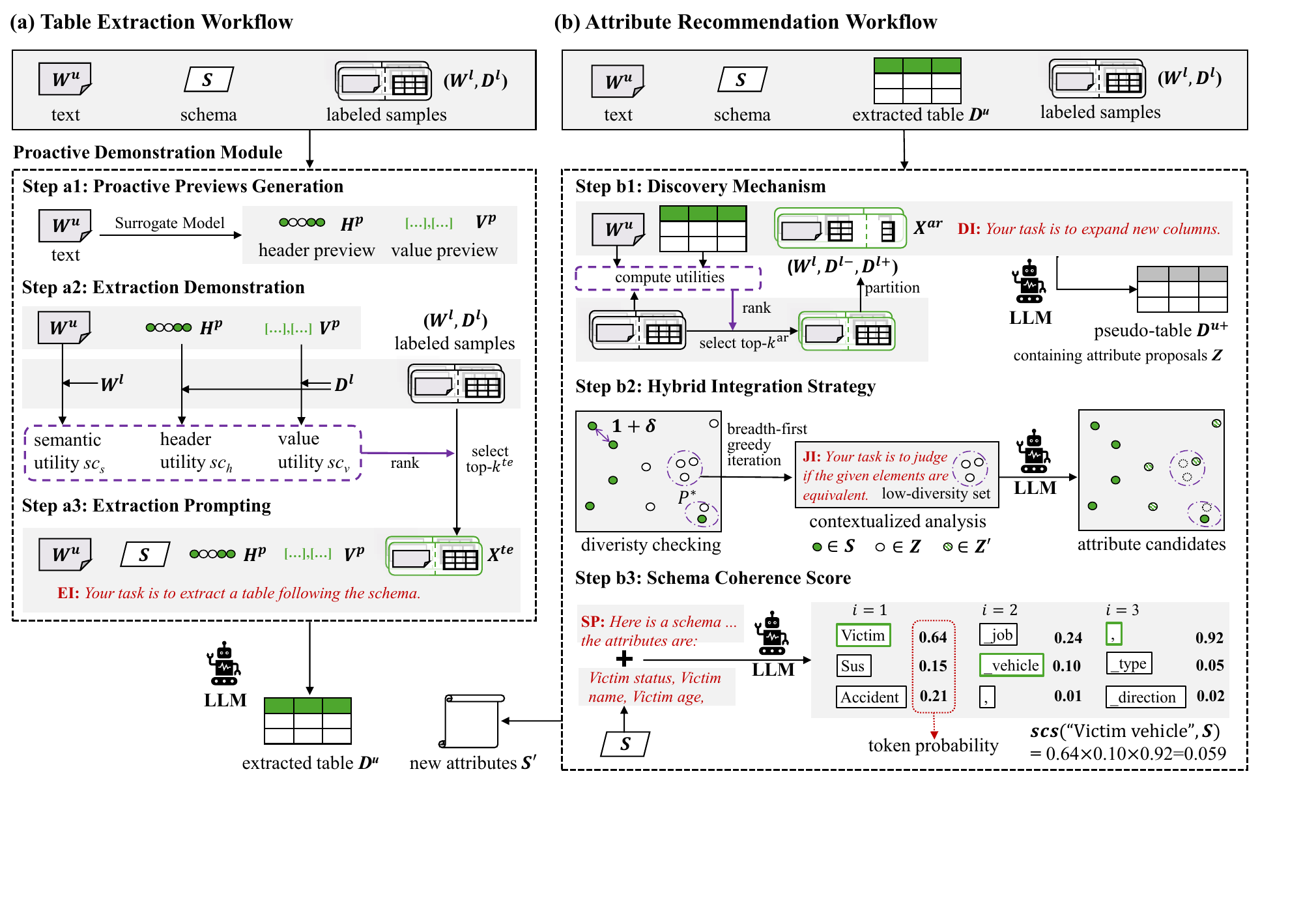}
    \caption{Overview of~\modelname. The LLM instructions are simplified, with the complete version in our published code.}
    \label{fig:tear-framework}
\end{figure*}
Fig.~\ref{fig:tear-framework} provides an overview of our dual-workflow framework, which leverages the instruction-following ability of a backbone LLM to perform table extraction with attribute recommendation. 
\textbf{(a) Table Extraction Workflow} (TEW): 
Given an input text, the schema, and the labeled samples, we dynamically select demonstrative examples from the labeled samples and insert them into the prompt for adaptive instruction. 
(Step a1) We employ a surrogate model to make fuzzy predictions about the target table, generating a header preview and a value preview.
(Step a2) For each labeled sample, we compute three utility scores by comparing its text to the input text, its table headers to the header preview, and its table values to the value preview.
(Step a3) The most pertinent examples under each score are selected and prompted to the LLM together with the static instruction to obtain the output table.
\textbf{(b) Attribute Recommendation Workflow} (ARW): 
Our ARW revisits the text to recommend attributes that are not yet present in the extracted table but are closely related to the existing schema.
(Step b1) The LLM is instructed to expand the extracted table by adding new columns.
(Step b2) We collect the newly expanded column headers from different tables, $Z$, to deduplicate and consolidate them into a global attribute candidate set $Z'$.
(Step b3) For each candidate, we compute the semantic coherence of appending it to the heuristic schema to produce a ranked list.

\subsection{Table Extraction Workflow}
\label{sec:tew}
Recalling Example~\ref{eg:c1}, we posit that demonstrative examples should bridge the gap between the LLM's general knowledge and task specification.
Therefore, we propose a \textbf{Proactive Demonstration Module} that first proactively forecasts what contents are likely to appear in the target table, then uses these forecasts to query the labeled pool for examples that disambiguate their extraction. 
Specifically, it forecasts a \textit{header preview} indicating the likely attribute names, and a \textit{value preview} suggesting plausible textual spans in the non-header region. 
Since previews only highlight fuzzy cues to guide example selection, their generation is inherently easier and more noise-tolerant than producing a precise, complete table. 
We fine‑tune a lightweight \textit{surrogate model} \revision{for each dataset} to generate them, which demands far fewer resources than end-to-end training.

\subsubsection{Proactive previews generation}
For the input text $W^u$, we define its header preview $H^p\subseteq S$, and its value preview $V^p\subset \mathcal{P}(W^u)$. We use the labeled pool $\{(W^l, D^l)\}$ to construct the training data for the surrogate model $M$.
Specifically, we serialize the table $D^l$ into a sequence $Seq(D^l)$, and optimize $M$ to generate this sequence autoregressively from $W^l$ with the standard cross-entropy loss ~\cite{sutskever2014sequence}. 
To serialize a table, we introduce three special tokens: 
$\langle m\rangle$ separates spans in each value,  $\langle s\rangle$ separates cells in each row, and $\langle n\rangle$ separates rows. 
The sequence representation of a cell value $A=[a_1, a_2, ..., a_{|A|}]$ is: 
\begin{equation}
    Seq(A)=a_1 \langle m \rangle\ a_2\ \langle m\rangle\ ... \langle m\rangle\ a_{|A|}
    \label{seq-a}
\end{equation}
The header sequence $Seq(H^l)$,the $i$-th data row sequence, and the entire table sequence are:
\begin{equation}
\begin{aligned}
&Seq(H^l) = h^l_1\ \langle s\rangle \  h^l_2 \  \langle s\rangle ...\langle s\rangle \  h^l_{m} ,\quad
Seq(R^l_i) = Seq(A^l_{i1})\ \langle s\rangle \ Seq(A^l_{i2}) \ \langle s\rangle ... \langle s \rangle \  Seq(A^l_{im})     
\\
&Seq(D^l) = Seq(H^l)\ \langle n \rangle \  Seq(R^l_1) \ 
     \langle n\rangle \  Seq(R^l_2) \ \langle n\rangle\ ... \langle n\rangle \  Seq(R^l_{n}) \label{seq-d}    
\end{aligned}
\end{equation}

During inference, given a new input text $W^u$, we obtain the predicted table sequence $M(W^u)$. 
Instead of reconstructing the full table, we probe it to obtain only $H^p$ and $V^p$: 
for $H^p$, $M(W^u)$ is truncated at the first $\langle n \rangle$ and split by $\langle s \rangle$;
for $V^p$, the sequence after the first $\langle n \rangle$ is split by all special tokens.
This yields coarse but informative previews that guide the subsequent example retrieval.

\subsubsection{Extraction demonstration}
Given the input text $W^u$ and the generated previews $H^p$ and $V^p$, we dynamically retrieve examples from the labeled pool based on three utility measures: header utility $sc_h$, value utility $sc_v$, and semantic utility $sc_s$. 

\paragraph{\underline{Header utility}} 
Samples whose table headers are similar to $H^p$ illustrate how to structure and populate attributes for the target table. 
We define the header utility of a labeled sample whose table $D^l$ has headers $H^l$ as 
\begin{equation}
    sc_h(H^p, D^l)= |H^p \cap H^l|+|H^l \setminus H^p|/(|S|+1).
\label{header-sim}
\end{equation}
The first term prioritizes samples that directly contain the anticipated headers; the second term, always less than 1, breaks ties by favoring samples with more additional headers, which provide broader contextual information. 
If $H^p=\emptyset$, meaning no header is previewed, Equation~\ref{header-sim} reduces to $|H^l|/(|S|+1)$, defaulting to retrieving samples with richer headers.

\paragraph{\underline{Value utility}}  
Samples whose table value spans are similar to $V^p$ demonstrate how salient mentions or descriptive phrases are mapped into table cells.  
Let $V^l= \cup_{i\in[1..n],j\in[1..m]} A^l_{ij}$ be the union of all spans in the labeled table $D^l$. We define the value utility as 
\begin{equation}
    sc_v(V^p, D^l)=\text{F1}(V^p, V^l; \text{chrF}\beta), \quad \text{chrF}\beta= (1+\beta^2)\frac{\text{chrP}\cdot \text{chrR}}{\beta^2 \text{chrP}+ \text{chrR}},
    \label{value-sim}
\end{equation} where F1$(\cdot)$ is the standard F1-score (Equation~\ref{eq:f1}) between two sets and chrF$\beta$~\cite{popovic2015chrf, popovic2011morphemes} is a widely used metric that measures pattern matching between two strings, with chrP and chrR denoting character n-gram precision and recall. We set $\beta$ to the previous default~\cite{popovic2016chrf}.
If $V^p=\emptyset$, meaning no value span is previewed, we will fall back to treating the original text $W^u$ as a single long span in place of $V^p$ in Equation~\ref{value-sim}.

\paragraph{\underline{Semantic utility}} 
Samples whose texts are overall semantically similar to $W^u$ offer a comprehensive demonstration in style and content. We compute the semantic utility by the standard dense retrieval practice in RAG~\cite{fan2024survey}, where a deep embedding model~\cite{meng2024sfrembedding}, $Emb(\cdot)$, is applied for normalized vectorization:
\begin{equation}
sc_s(W^u, W^l) = Emb(W^u)^{\mathsf{T}}\cdot  Emb(W^l).
\end{equation}

\subsubsection{Extraction prompting} 
For simplicity, we retrieve an equal number of examples under each utility type.
The number $k^{te}$ is a hyperparameter affected by the backbone LLM's capability and can be tuned via validation performance.
Let $\mathcal{X}_h$, $\mathcal{X}_v$, $\mathcal{X}_s$ be the top-$k^{te}$ samples indices according to $sc_h$, $sc_v$, $sc_s$, respectively. 
The \textbf{Proactive Demonstration Module} outputs demonstration set: \begin{equation}
    X^{te}=\{(W^l_i, D^l_i)\}_{i\in \mathcal{X}_h\cup \mathcal{X}_v \cup \mathcal{X}_s }.
\end{equation} Here it is reasonable to use the same pool for training $M$ and as the retrievable set, because at the retrieval is invisible to the already trained model. 
Splitting the pool only shrink both sets to affects the performance under data scarcity (Section~\ref{sec:ablation-pool}). 
We then use a natural language template with reasonable instructional phrases and formatting markers to wrap the text, schema, auxiliary previews, and retrieved demonstrations into a single prompt, the Extraction Instruction (EI).
Fig.~\ref{fig:tear-framework} shows a conceptualized template, while a complete instantiation is in our code.
The final table extraction is then performed as
\begin{equation}
    D^u = LLM(\text{EI}(W^u, S, H^p, V^p, X^{te})).
\end{equation} 
\revision{
The above emphasizes \emph{what} to populate the template, instead of the particular languages of EI.}

\subsection{Attribute Recommendation Workflow}
\label{sec:arw}
We formulate the Attribute Recommendation (AR) task to support the emerging exploratory scenarios for n‑texts, where the user provides only a heuristic schema as an initial direction, and the system is responsible for interacting with the texts to return a ranked list of new attributes. 
\revision{
Importantly, AR does not produce a conclusive attribute set, but rather exhibits a prioritized list serving as an automatic, intelligent summary of text-driven attributes.
}
We have identified three essential abilities that an effective LLM-based AR method should possess in the Introduction, and our ARW instantiates corresponding components, as depicted in Fig.~\ref{fig:tear-framework}(b). 
Briefly, it comprises three core components: a \textbf{Discovery Mechanism} that proposes new attributes from input texts via adaptive demonstration, a \textbf{Hybrid Integration Strategy} that efficiently resolves duplicates among generated proposals, and a \textbf{Schema Coherence Score} that ranks candidates by their relevance to the user's initial schema.
We detail each component in the following subsections.

\subsubsection{Discovery Mechanism}
\label{sec:dm}
To discover new attributes from an input text, we continue the idea of adaptive instruction, demonstrating the desired behavior through examples.
Our Discovery Mechanism places attributes back into their naive context as column headers, and guides the LLM to expand the extracted table with new columns by drawing analogies to existing ones. 
By framing open‑ended discovery as table expansion, we ground the LLM's understanding of attributes in the structure of table columns, thereby preserving their original semantics more faithfully.

The input to the mechanism is the text $W^u$ and the extracted table $D^u$, and the output is a \textit{pseudo-table} $D^{u+}$ which contains columns as new attribute proposals. 
A demonstrative example for this step takes the form (text $W^l$, known table $D^{l-}$, new table $D^{l+}$), which is constructed by splitting the columns of the sample table part $D^l$.
\revision{\begin{example}
Suppose a labeled sample $D^l$ has columns: ``Victim name'', ``Victim status''. To create a discovery example for an input text whose $D^u$ has only a column ``Victim name'', we split $D^l$ into $D^{l-}$ of column ``Victim name'', and $D^{l+}$ of ``Victim status''. 
\end{example}}
Note that although the attributes in the demonstrative “new" table belong to the existing schema $S$, we do not reveal $S$ to the LLM. 
This allows us to reuse the same labeled pool originally collected for $S$ to effectively demonstrate an open-ended discovery task.
When retrieving the examples, we also update the previews with the header and value views derived from $D^u$. Let $\tilde{\mathcal{X}}_h$ and $\tilde{\mathcal{X}}_v$ be the indices of the updated retrieved samples. The discovery demonstration set $X^{ar}$ and the pseudo-table extraction are: 
\begin{equation}
    X^{ar}=\{(W^l_i, D^{l-}_i, D^{l+}_i)\}_{i\in \tilde{\mathcal{X}}_h \cup \tilde{\mathcal{X}}_v \cup \mathcal{X}_s}, \quad   D^{u+}=LLM(\text{DI}(W^u, D^u, X^{ar}) ),
\end{equation} where the Discovery Instruction (DI) is like EI.

\subsubsection{Hybrid Integration Strategy}
\label{sec:his}
There can be duplicates among individual discoveries.
For instance, conceptually equivalent attributes may appear as "Hospital name" in one pseudo‑table and "Medical center name" in another. 
LLMs are adept at detecting such duplicates through semantic reasoning over attribute names and their textual contexts.~\cite{witteveen2019paraphrasing, dai2025auggpt}.
We solicit a binary judgement $\{\textsf{True}, \textsf{False}\}$ from the LLM over a small set of proposals at each time. 
This constrained format could effectively prevent the LLM from drifting into verbose or open‑ended reasoning~\cite{liu2024lost, peng2023does}, thereby preserving controllability.

The input of the strategy are schema $S$ and proposals $Z=\cup_{i} H^{u+}_{i}\setminus S$, and the output is a deduplicated subset $Z'\subseteq Z$.
As $Z$ grows, efficiency further draws our attention, as plainly invoking the LLM on all possible combinations becomes prohibitive.
Therefore, we first introduce a lightweight \textit{diversity checking} to identify a few plausible proposal subsets, and only submit those low-diversity subsets to the LLM for \textit{contextualized analysis}.

\paragraph{\underline{Diversity checking}}
For a group of attributes $ p\subseteq Z\cup S$, we compute its diversity as the maximum Vendi Score~\cite{dan2023vendi, mironovmeasuring} over its subset:
\begin{equation}
    div(p)=\max_{q\subseteq p}  vds(q),
    \label{eq:div}
\end{equation} where $vds(q)\in [1, |q|]$ is a continuous number that estimates the effective number of unique elements in $q$.
%
The collection of low-diversity subsets $P^*$ is defined as:
\begin{equation}
    P^* = \{p\subseteq Z \cup S | |p|>1, div(p)<1+\delta\}, \quad 
    1+\delta = \min_{\forall s_i,s_j\in S, s_i\neq s_j} div(\{s_i, s_j\}).
\label{eq:suspecious}
\end{equation}
Here, $1+\delta$ is the minimum diversity between any two distinct schema attributes.
If $div(p)<1+\delta$, $p$ approximates only a single attribute under the schema's semantic granularity, and thus may accommodate duplicates. 
We could use a bottom-up search with pruning to find $P^*$ given that $div(\cdot )$ is monotonic~\cite{mironovmeasuring}(see Appendix~\ref{app:p-star}).

\paragraph{\underline{Contextualized analysis}}
For each proposal $z$ involved in some low-diversity subset, we obtain its context$(z)$ by prompting the LLM to produce a concise definition~\cite{zhang2024extract}, based on its native texts and pseudo-tables, so that contexts share a consistent style and length and
can be compared fairly.
%
Then, for each $p\in P^*$, a Judge Instruction (JI) is prompted to the LLM, 
\begin{equation}
 rsp = LLM(\text{JI}(S, \{(z, \text{context}(z))\}_{z\in p})), \  rsp\in \{\textsf{True}, \textsf{False}\}. \label{eq:ji}
\end{equation}
If the LLM responds that $p$ is duplicated, then we check whether $p$ contains a known attribute or not. 
If so, all other proposals are deemed naive repetitions of that known attribute and will be eliminated. If not, we choose the proposal in $p$ with the highest Schema Coherence Score (Section~\ref{sec:rlv}) and discard the rest.

We could use earlier judgment to remove some proposals, so not every $p\in P^*$ requires contextualized analysis. We hence design a breadth‑first greedy iteration scheduling that dynamically updates $P^*$ to reduce LLM calls.
As in Algorithm~\ref{alg:bfs}, each outer iteration selects a cover $Q\subset P^*$ of all remaining elements, and the inner loop sequentially submits $Q$ to the LLM. 
This breadth‑first design prunes confirmed duplicates early, shrinking $P^*$ and avoiding unnecessary exploration of nested sets that do not contain genuine duplicates.

We now briefly summarize the computational complexity of the Hybrid Integration Strategy, and refer to Appendix \ref{app:p-star} for detailed discussions and the Table~\ref{tab:eff} for the experiment report.
In diversity checking, the number of candidate sets examined by a bottom-up search scales linearly with the output size $|P^*|$. 
In contextualized analysis, let $|B|$ be the number of elements at the start of an iteration, and $N'$ be the number of true duplicates. The number of LLM calls in that iteration is bounded by $(\ln |B|+1)(|B|-N')$. More duplicates yield a smaller bound and likely earlier \textsf{True} responses for more aggressive pruning. This property is beneficial: when duplicates are abundant, the algorithm eliminates large groups with few calls; when scarce, it degrades gracefully to exhaustive iteration. Even when the resource cannot afford the exhaustive iteration, early termination incurs little penalty since there are fewer duplicates in the first place.


\begin{algorithm}[t]
\caption{Breadth-First Greedy Iteration of $P^*$}\label{alg:bfs}
\begin{algorithmic}[1]
\Require attribute proposals $Z$, schema $S$, low-diversity set $P^*$
\Ensure attribute candidates $Z'$
\State $Z'=\emptyset$, $B \leftarrow Z\cup S$
\While{$|B|>0 \land |P^*|>0$} \Comment{Outer loop}
   \State $B'\leftarrow B, Q\leftarrow\emptyset $
   \While{$|B'|>0$ } \Comment{Find a greedy cover of elements.}
    \State $q=\arg \max_{q\in P^*} q\cap B' $  
    \State $Q\leftarrow Q\cup\{q\}, B'\leftarrow B'\setminus q$
   \EndWhile
    
    \For{$q'\in Q$} \Comment{Inner loop}
        \State $q\leftarrow q' \cap B$ 
        \If{$|q|=1$} $Z' \leftarrow Z'\cup q$ 
        \Else
            \State Obtain $rsp$ for $q$ by Equation~\ref{eq:ji}.
            \If{$rsp$}
                \If{$q\cap S=\emptyset$} $Z' \leftarrow Z'\cup \{\arg\max_{z\in q} scs(z, S)\}$
                \EndIf  
                \State $B\leftarrow B\setminus q$, $P^*\leftarrow \{p \in P^*|p \cap q= \emptyset\}$  \Comment{Exclude elements and sets.}
            \Else \ $P^*\leftarrow P^*\setminus\{q\}$
            \EndIf
        \EndIf
    \EndFor
\EndWhile
\State $Z'\leftarrow Z'\setminus S$
\end{algorithmic}
\end{algorithm}

\subsubsection{Schema Coherence Score}
\label{sec:rlv}
It is necessary to prioritize attributes that meaningfully extend the user's heuristic schema.
For example, while ``Reporter name'' may be informative in a news article, it adds little value when the schema is designed to extract facts about criminal incidents rather than newsroom staffing.
To quantify this notion of relevance, we introduce a Schema Coherence Score, which estimates how naturally an attribute candidate completes the description of the existing schema.
Based on the language modeling principles~\cite{jelinek1977perplexity,petroni2019language, sileo2022zero}, we compute this score as the conditional probability of the candidate's name given the known schema.
This formulation captures holistic coherence with the schema context.

Formally, let $z$ be a candidate. We tokenize $z$ and append an end‑of‑name separator $b_0$ (a comma by default):
\begin{equation}
    tkn(z)= Tokenizer(z)+Tokenizer(b_0).
\end{equation} 
We then prepend a Schema Prefix (SP) that enumerates the known attributes as the condition. As in Fig.~\ref{fig:tear-framework} (Step b3), the score is
\begin{equation}
    scs(z, S)=\Pi_i  Prob(tkn_i(z)|\text{SP}(S)+tkn_{:i-1}(z)), 
    \label{eq:rlv}
\end{equation} where $tkn_i(z)$ is the $i$-th token and $tkn_{:i-1}(z)$ denote the tokens before it.
Computing this score requires no decoding, only a single forward pass through the LLM, making it substantially more efficient than generating language responses.
Eventually, ARW ranks each $z\in Z'$ by $scs(z,S)$ and presents the top‑$k$ to the user.
\section{Evaluation}
\label{sec:eval}
\begin{table*}[t]
\centering
\caption{Benchmark statistics. 1, 2, and 3 in the header region are exploratory levels. }
\footnotesize
\begin{tabular}{l|ccccccc|ccccccc}
\toprule
 &  & \multicolumn{2}{c}{\#Token} & \multicolumn{2}{c}{\#Record} & \multicolumn{2}{c}{\#Column} &  & \multicolumn{3}{c}{$|S|$} & \multicolumn{3}{c}{$|\hat{S}'|$} \\
\cmidrule(lr){3-4} \cmidrule(lr){5-6} \cmidrule(lr){7-8} \cmidrule(lr){10-12} \cmidrule(lr){13-15}
Datasets &\#Text & Avg. & Max. & Avg. & Max. & Avg. & Max. & \#Attr.& 1 & 2 &  3 &  1 &  2 &  3 \\
\midrule
Incidents & 1,369 & 24.1 & 97 & 1.6 & 6 & 3.1 & 10 & 28 & 22 & 18 & 14 & 6 & 10 & 14 \\
Weather & 2,006 & 22.3 & 95 & 1.3 & 5 & 2.9 & 7 & 19 & 15 & 12 & 9 & 4 & 7 & 10 \\
\revision{Conversation} & \revision{1,000} & \revision{303} & \revision{938} & \revision{1} & \revision{1} & \revision{8.3} & \revision{24} &\revision{35} & \revision{28} &\revision{23} & \revision{18}& \revision{7} &\revision{12}  & \revision{17} \\
\bottomrule
\end{tabular}
\label{tab:stats}
\end{table*}
Given that previous evaluation methodologies are unsuitable for our focus, we introduce new datasets and applicable metrics.

\subsection{Datasets and Splits}
The input texts of existing table extraction datasets are table descriptions or specialized documents that cannot reflect the distribution of n-texts. 
To verify our focus, we present two real‑world datasets, which together provide a total of 3,375 (text, table) pairs:
 \underline{\emph{Incidents}}. The texts are gun‑violence news reports, and tables capture victims, suspects, and accidents, with attributes such as “Victim name”, “Suspect name”, and “Accident address”.
 \underline{\emph{Weather}}. The texts are weather forecasts for multiple countries, and tables include attributes such as “Weather frequency” and “Wind speed”.
Both texts are news reports scraped by the CACAPO project~\cite{van2020cacapo}, gathered without any extraction task in mind, and therefore exhibit realistic linguistic variability. 
\revision{
The ground truth schema is derived from human responses to the 5W1H aspects (who, what, when, etc.) of each text, reflecting genuine text-driven needs instead of being arbitrarily carved.
}
The original release contains some attribute annotations but lacks complete tables (values are not aligned to form multiple records). We hired university students to annotate complete tables while correcting errors (Appendix~\ref{app:data}). 

Besides the news datasets, we further adapt existing conversational texts, MultiWoz2.4~\cite{multiwoz}, to our focus:  \underline{\emph{Conversation}}. The texts are multi-topic, multi-turn human-human dialogues around services such as attractions, hotels, and restaurants. %
The dialogue states cover salient contents and can be converted to text-driven attributes, whose values may change as people change their requests or make clarifications. For table extraction, we sample 1000 texts and set labels as the final agreed-upon attribute values, while value drifts introduce intrinsic semantic noise.
We use Conversation as a controlled stress test on verbosity and noise that does not diminish the contribution of the new datasets. The reason is that the dialogues have natural language patterns yet are oriented to a closed service ontology; thus, it is less variable than the new datasets, of which a statistical examination is in Appendix~\ref{app:variability}. 

Table~\ref{tab:stats} summarizes the statistics, and the datasets are in our codebase. 
To reflect the application need under data scarcity, we randomly reserve 300 samples as the labeled set \revision{for all methods}.
For TE, we report performance using the full schema (Section~\ref{sec:te-exp}). 
For AR, we further create three exploratory levels: we first identify low‑frequency attributes that appear in <15\% of samples, which non‑experts would likely miss during heuristic schema design. We then randomly drop such attributes so that the heuristic schema lacks 20\%, 35\%, or 50\% of all attributes. 
The absent attributes are completely hidden from the system, and the system is also unaware of the exploratory level (Section~\ref{sec:ar-exp}). 
We also evaluate the extracted table under these exploratory settings, comparing it against the table following the full schema (Section~\ref{sec:exp-tear}). 

\begin{figure}[t]
    \centering
    \includegraphics[trim=5 455 450 0, clip, width=0.6\linewidth]{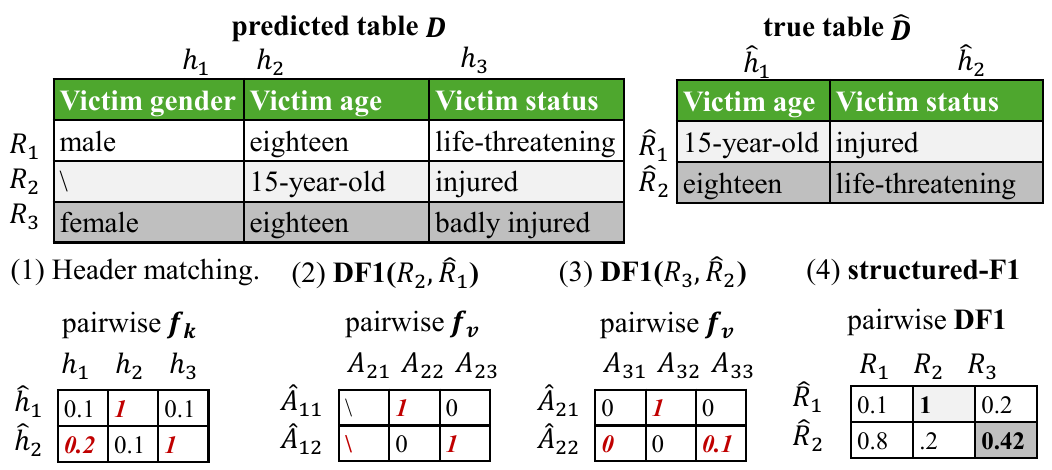}
    \caption{Evaluation example.}
    \label{fig:metric}
\end{figure}
\subsection{Metrics}
\subsubsection{Table extraction metrics}
Let $D$ be the predicted table with headers $H$ and records $\mathcal{R}$, and $\hat{D}$ be ground truth with $\hat{H}$ and $\hat{\mathcal{R}}$. 
We introduce table extraction metrics with the example in Figure~\ref{fig:metric}.

\paragraph{\underline{header-F1}}
Previous work use header-F1=F1($H,\hat{H}; f$) to evaluate whether the system correctly identifies the attributes mentioned in the text. 
Here, F1 is the standard F1-score between two sets, and $f$ is a given similarity function for comparing two elements (e.g., exact match, or soft string similarity). Formally, 
\begin{equation}
    \begin{aligned}
\text{P}(X,Y; f)&=\frac{1}{|X|}\sum_{x\in X} \max_{y\in Y} f(x,y),  \quad 
\text{R}(X,Y; f)=\frac{1}{|Y|}\sum_{y\in Y} \max_{x\in X}f(x,y),\\
\text{F1}(X,Y;f)&=2/(\text{P}(X,Y; f)^{-1}+\text{R}(X,Y;f)^{-1}).
    \end{aligned}
    \label{eq:f1}
\end{equation} 
\begin{example}
The header regions are $\hat{H}$=\{Victim age, Victim status\}, $H$=\{Victim gender, Victim age, Victim status\}. Given extract match $f$, P$(H,\hat{H}; f)$=0.67, R$(H,\hat{H}; f)$=1, header-F1=F1$(H,\hat{H}; f)$=0.8.
\end{example}

\paragraph{\underline{structured-F1}} 
To evaluate whether the values are correctly extracted and aligned into a table, 
existing works largely follow the approach of Wu et al.~\cite{wu2022text}, flattening a table into (header, index, value) triples. 
For instance, the table in Figure~\ref{fig:source-text}(a) uses team names as the logical index, yielding triples such as “(Losses, Hawks, 12)”. 
However, n‑texts rarely provide a natural index column, and forcing one arbitrarily distorts the table's native semantics.
We therefore adopt a more fundamental view of a table: a \textit{set of records}, where each record is a self‑contained header‑to‑value dictionary.
Under this view, comparing a predicted table to the ground truth reduces to matching two sets, for which standard F1 naturally solves.
Crucially, this formulation respects the integrity of each record, and alignment is resolved through record matching, rather than an externally imposed index.  
We thus propose \textbf{structured-F1}:
\begin{equation}
\text{structured-F1=F1(}\mathcal{R}, \hat{\mathcal{R}}\text{; DF1)},
\end{equation}where DF1 (Dictionary F1) measures the similarity between two records. 
Specifically, each record $R_i$ is a dictionary mapping headers to values, with domain $\operatorname{dom}(R_i)=H$ and $R_i(h_j)=A_{ij}$. 
DF1 computes the similarity between $R_i$ and $\hat{R}_j$ in two steps: it first aligns headers using a similarity function $f_k$, then aggregates the corresponding value similarities via $f_v$.
Formally,
\begin{equation}
\begin{aligned}
&\text{DP}(R_i, \hat{R}_j;f_k,f_v) = \frac{1}{|R_i|} \sum_{h \in H} f_v\Bigl( R_i(h), \hat{R}_j(\arg \max_{\hat{h} \in \hat{H}} f_k(h, \hat{h})) \Bigr), \\
&\text{DR}(R_i, \hat{R}_j;f_k,f_v) = \frac{1}{|\hat{R}_j|} \sum_{\hat{h} \in \hat{H}} f_v\Bigl( R_i(\arg\max_{h \in H} f_k(h, \hat{h})), \hat{R}_j(\hat{h}) \Bigr), \\
&\text{DF1}(R_i, \hat{R}_j;f_k,f_v) = 2/(\text{DP}(R_i, \hat{R}_j;f_k,f_v)^{-1} + \text{DR}(R_i, \hat{R}_j;f_k,f_v)^{-1}).
\end{aligned}
\end{equation}

\begin{example}
(1) Compute the matched header to be invariant to the column order. (2) The matched position (bolded) is applied to aggregate the value similarity. For $R_2$ and $\hat{R}_1$, the empty value is ignored: DP=1, DR=1, and DF1$(R_2, \hat R_1)$=1. (3) For $R_3$ and $\hat{R}_2$, DP=(0+1+0.1)/3=0.37, DR=(1+0.1)/2=0.55, so DF1$(R_3, \hat R_2)$=0.42. 
(4) After computing for all 6 record pairs, P($\mathcal{R}$, $\mathcal{\hat{R}}$)=(0.8+1+0.42)/3=0.74, R($\mathcal{R}$, $\mathcal{\hat{R}}$)=(1+0.42)/2=0.71, structured-F1=0.73.
\end{example}

The complexity of structured-F1 is not higher than the previous evaluation (Appendix~\ref{app: f1}). We use the same three string similarities as existing works: Extract Match (EM), chrF$\beta$ (chrf)~\cite{popovic2015chrf}, and rescaled BERTScore (BS) ~\cite{zhangbertscore}. For values with multiple spans indicating multiple mentions, the string similarity is first computed span-wise and aggregated via F1. 

\subsubsection{Attribute Recommendation Metrics}  
We evaluate AR quality using recall, as is standard in recommendation tasks.
Let $S'=Z'[:k]$ denote the top‑$k$ attribute candidates.
To aggregate performance across different $k$, we compute the recall curve $\text{R}(Z'[:k], \hat{S}', f)$ as a function of $k$, and report the Area Under the Curve (recall‑AUC) as the overall metric.
When comparing curves of different lengths, shorter curves are padded with their final value to ensure consistent normalization.
A strong recommendation list reaches higher recall earlier, leading to a larger recall‑AUC.
For the similarity function $f$, we use only chrf and BS because asking for an exact match under open‑ended discovery is overly stringent for practical use.

\section{Experiments}
\label{sec:exp}
\begin{table*}[th]
 \caption{Table extraction performance (\%). \revision{The best results within each backbone are \textbf{bolded}, and the best in the same row are \underline{underlined}.}}
  \label{tab:te}
   \footnotesize
\resizebox{\textwidth}{!}{
\begin{tabular}{c|ll|c|ccc|ccc|ccc}
    \toprule
    &  & &  & \multicolumn{3}{c}{Llama-3-8B-Instruct} & \multicolumn{3}{c}{Qwen2.5-14B-Instruct} & \multicolumn{3}{c}{\revision{Llama-3-70B-Instruct}} \\
    Dataset & Metric & sim. & TRE & MapMake & RAG & \modelname & MapMake & RAG & \modelname & \revision{MapMake} & \revision{RAG} & \revision{\modelname} \\
    \midrule
 \multirow{6}{*}{\shortstack{Inci-\\dents}}&  \multirow{3}{*}{\shortstack{header-\\F1}} &EM & 86.7$\pm$0.8 
  & 47.4$\pm$0.8 
  & 81.5$\pm$0.1 
  & \textbf{87.8}$\pm$0.3 
  & 68.1$\pm$0.3 
  & 83.3$\pm$0.1 
  & \textbf{89.1}$\pm$0.4
& \revision{68.1$\pm$0.2} 
& \revision{86.8$\pm$0.0} 
& \revision{\underline{\textbf{90.3}}$\pm$0.1} \\

& & chrf 
  & 91.2$\pm$0.7 
  & 54.8$\pm$0.9 
  & 87.9$\pm$0.1 
  & \textbf{92.5}$\pm$0.3 
  & 74.9$\pm$0.2 
  & 88.2$\pm$0.0 
  & \textbf{92.8}$\pm$0.2 
  & \revision{74.3$\pm$0.3}
  & \revision{91.6$\pm$0.0} 
  & \revision{\underline{\textbf{93.9}}$\pm$0.1} \\

& & BS   
  & 91.9$\pm$0.7 
  & 55.3$\pm$0.9 
  & 88.5$\pm$0.1 
  & \textbf{93.1}$\pm$0.3 
  & 75.8$\pm$0.2 
  & 88.9$\pm$0.0 
  & \textbf{93.4}$\pm$0.2 
  & \revision{75.0$\pm$0.1} 
  & \revision{92.3$\pm$0.0} 
  & \revision{\underline{\textbf{94.5}}$\pm$0.1} \\

\cmidrule{2-13}
&  \multirow{3}{*}{  \shortstack{struc-\\tured-\\F1}} &EM 
  & 64.5$\pm$0.1 
  & 25.7$\pm$0.4 
  & 61.2$\pm$0.2 
  & \textbf{69.1}$\pm$0.2 
  & 40.9$\pm$0.4 
  & 62.0$\pm$0.0 
  & \textbf{70.6}$\pm$0.1 
   & \revision{44.8$\pm$0.3} 
  & \revision{69.8$\pm$0.0} 
  & \revision{\underline{\textbf{73.3}}$\pm$0.0} \\

& & chrf 
  & 74.9$\pm$0.5 
  & 34.1$\pm$0.3 
  & 73.6$\pm$0.1 
  & \textbf{79.4}$\pm$0.1 
  & 54.1$\pm$0.2 
  & 76.8$\pm$0.0 
  & \textbf{82.0}$\pm$0.2
& \revision{56.5$\pm$0.1} 
  & \revision{80.0$\pm$0.0} 
  & \revision{\underline{\textbf{82.4}}$\pm$0.1}\\

& & BS   
  & 78.8$\pm$0.4 
  & 42.5$\pm$0.2 
  & 76.8$\pm$0.1 
  & \textbf{82.9}$\pm$0.1 
  & 60.5$\pm$0.4 
  & 74.9$\pm$0.0 
  & \textbf{82.4}$\pm$0.2
  & \revision{61.9$\pm$0.2} 
  & \revision{81.5$\pm$0.0} 
  & \revision{\underline{\textbf{84.6}}$\pm$0.1} \\
      \midrule
\multirow{6}{*}{\shortstack{Wea-\\ther}}&  
 \multirow{3}{*}{\shortstack{header-\\F1}} &EM & 79.1$\pm$0.5 
  & 52.2$\pm$1.3 
  & 77.3$\pm$0.0 
  & \textbf{80.3}$\pm$0.4 
  & 66.1$\pm$0.0 
  & 80.4$\pm$0.1 
  & \textbf{82.6}$\pm$0.2 
 & \revision{74.3$\pm$0.0} 
  & \revision{83.9$\pm$0.1} 
  & \revision{\underline{\textbf{84.1}}$\pm$0.1} \\

& & chrf 
  & 84.9$\pm$0.3 
  & 58.9$\pm$1.3 
  & 83.5$\pm$0.0 
  & \textbf{85.6}$\pm$0.1 
  & 73.8$\pm$0.1 
  & 85.7$\pm$0.1 
  & \textbf{87.2}$\pm$0.1 
  & \revision{80.1$\pm$0.2} 
  & \revision{88.0$\pm$0.1}
  & \revision{\underline{\textbf{88.4}}$\pm$0.1} \\
& & BS   
  & 89.6$\pm$0.3 
  & 68.3$\pm$0.9 
  & 88.7$\pm$0.0 
  & \textbf{90.1}$\pm$0.0 
  & 82.1$\pm$0.0 
  & 89.9$\pm$0.1 
  & \textbf{91.3}$\pm$0.1 
  & \revision{86.7$\pm$0.0} 
  & \revision{91.9$\pm$0.1} 
  & \revision{\underline{\textbf{92.2}}$\pm$0.1} \\
\cmidrule{2-13}
&  \multirow{3}{*}{ \shortstack{struc-\\tured-\\F1}} &EM & 44.1$\pm$0.5 
  & 25.6$\pm$0.4 
  & 46.2$\pm$0.1 
  & \textbf{51.0}$\pm$0.4 
  & 39.3$\pm$0.3 
  & 50.7$\pm$0.1 
  & \textbf{53.9}$\pm$0.4 
 & \revision{43.2$\pm$0.3} 
  & \revision{54.9$\pm$0.1} 
  & \revision{\underline{\textbf{56.3}}$\pm$0.1} \\

& & chrf 
  & 63.7$\pm$0.3 
  & 42.1$\pm$0.8 
  & 65.4$\pm$0.1 
  & \textbf{70.1}$\pm$0.2 
  & 62.0$\pm$0.3 
  & 71.3$\pm$0.0 
  & \underline{\textbf{73.4}}$\pm$0.2 
  & \revision{64.5$\pm$0.2} 
  & \revision{72.1$\pm$0.1} 
  & \revision{\textbf{73.3}$\pm$0.0} \\

& & BS   
  & 63.0$\pm$0.3 
  & 42.2$\pm$1.3 
  & 64.3$\pm$0.1 
  & \textbf{67.9}$\pm$0.4 
  & 57.2$\pm$0.1 
  & 66.4$\pm$0.0 
  & \textbf{69.1}$\pm$0.5 
 & \revision{62.5$\pm$0.2} 
  & \revision{70.2$\pm$0.1} 
  & \revision{\underline{\textbf{71.5}}$\pm$0.0}\\
  \midrule
     \multirow{6}{*}{\revision{\shortstack{Conver-\\sation}}}&  \multirow{3}{*}{\revision{\shortstack{header-\\F1}}} & \revision{EM} & \revision{90.0$\pm$0.5} & \revision{52.7$\pm$2.0} & \revision{84.6$\pm$0.1} & \revision{\textbf{88.7}$\pm$0.2} & \revision{76.3$\pm$0.1} & \revision{87.4$\pm$0.0} & \revision{\underline{\textbf{90.7}}$\pm$0.1} & \revision{78.9$\pm$0.0} & \revision{86.8$\pm$0.0} & \revision{\textbf{90.1}$\pm$0.1} \\
& & \revision{chrf} & \revision{94.7$\pm$0.3} & \revision{62.2$\pm$1.9} & \revision{91.9$\pm$0.1} & \revision{\textbf{94.5}$\pm$0.1} & \revision{85.3$\pm$0.2} & \revision{93.4$\pm$0.0} & \revision{\underline{\textbf{95.3}}$\pm$0.0} & \revision{86.3$\pm$0.1} & \revision{93.4$\pm$0.0} & \revision{\textbf{95.1}$\pm$0.1} \\
& & \revision{BS} & \revision{94.3$\pm$0.3} & \revision{61.5$\pm$1.9} & \revision{91.3$\pm$0.1} & \revision{\textbf{93.9}$\pm$0.0} & \revision{84.4$\pm$0.1} & \revision{92.8$\pm$0.0} & \revision{\underline{\textbf{94.9}}$\pm$0.1} & \revision{85.9$\pm$0.1} & \revision{92.9$\pm$0.0} & \revision{\textbf{94.6}$\pm$0.1} \\
\cmidrule{2-13}
&  \multirow{3}{*}{\revision{\shortstack{struc-\\tured-\\F1}}} & \revision{EM} & \revision{80.9$\pm$0.6} & \revision{44.4$\pm$1.9} & \revision{76.9$\pm$0.1} & \revision{\textbf{82.4}$\pm$0.1} & \revision{66.2$\pm$0.1} & \revision{81.1$\pm$0.0} & \revision{\underline{\textbf{85.1}}$\pm$0.2} & \revision{69.4$\pm$0.0} & \revision{80.8$\pm$0.0} & \revision{\textbf{84.6}$\pm$0.1} \\
& & \revision{chrf} & \revision{85.4$\pm$0.5} & \revision{49.3$\pm$1.9} & \revision{81.2$\pm$0.1} & \revision{\textbf{86.1}$\pm$0.1} & \revision{71.5$\pm$0.1} & \revision{85.1$\pm$0.0} & \revision{\underline{\textbf{88.3}}$\pm$0.2} & \revision{74.8$\pm$0.1} & \revision{84.6$\pm$0.0} & \revision{\textbf{87.8}$\pm$0.1} \\
& & \revision{BS} & \revision{87.0$\pm$0.3} & \revision{53.6$\pm$2.1} & \revision{83.8$\pm$0.2} & \revision{\textbf{88.4}$\pm$0.1} & \revision{75.8$\pm$0.2} & \revision{87.6$\pm$0.0} & \revision{\underline{\textbf{90.3}}$\pm$0.1} & \revision{77.8$\pm$0.0} & \revision{86.7$\pm$0.0} & \revision{\textbf{89.7}$\pm$0.2} \\
\bottomrule
  \end{tabular}
}
\end{table*}

Our experiments answer the following questions about \modelname: \textbf{Q1}: How is the table extraction performance? \textbf{Q2}: How is the attribute recommendation performance? \textbf{Q3}: How do text-driven new attributes benefit table extraction? \textbf{Q4}: How is the efficiency? \textbf{Q5}: How are the ablation results?

\subsection{Setup}

\subsubsection{Baselines}
For table extraction, we compare baselines from both the supervised and the ICL paradigms. Note that some other ICL approaches rely on knowledge about the specialized documents (e.g., external KGs~\cite{jiang2024tkgt} or type recognizing s~\cite{jiang2025tst}) to decompose extraction into subtasks and design dedicated instructions, making them difficult to apply to n-texts.
1. \textbf{TRE}~\cite{wu2022text} is one of the best supervised methods, which augments the sequence generation model with special table relation embeddings.\footnote{https://github.com/shirley-wu/text\_to\_table}.
2. \textbf{MapMake}~\cite{ahuja2025map} is a recent ICL method. We implement the one‑shot version by selecting the labeled sample with the most headers and carefully writing the required reasoning steps.
3. \textbf{RAG}~\cite{zhang2025retrieval, meng2024sfrembedding}. We implement a strong baseline of dense RAG that could also dynamically choose examples to contrast with our Proactive Demonstration Module.

As for the novel attribute recommendation task, we compare against three strong baselines that all prompt the LLM with retrieved examples to obtain attribute proposals, and then adopt different integration and ranking strategies. The comparison does not include a fully independent end-to-end attribute-discovery baseline, since we did not find a stable and decent alternative for our Discovery Mechanism.
4. \textbf{Corpus Frequency Ranking} (CFR) integrates and ranks attribute proposals according to their global frequency in the corpus. The intuition is that attributes mentioned more frequently across texts are more important.
5. \textbf{Maximum Schema Similarity} (MSS) ranks proposals by their maximum semantic similarity to the known schema attributes, $sim(z,S)$ = max$_{s\in S} Emb($context$(z))\cdot Emb($context$(s))$. The motivation is that newly discovered attributes should be semantically closer to those already known.
6. \textbf{Direct LLM Reasoning} (DIRECT) prompts an LLM to rank the proposals as the most suitable for extending the known schema, considering relevance and usefulness.

\subsubsection{Implementation}
For all compared methods, we employ three instruction-tuned open-source LLMs: Llama-3-8B-Instruct\footnote{\url{https://huggingface.co/meta-llama/Meta-Llama-3-8B-Instruct}}, Qwen2.5-14B-Instruct\footnote{\url{https://huggingface.co/Qwen/Qwen2.5-14B-Instruct}}, and \revision{Llama-3-70B-Instruct}\footnote{\url{https://huggingface.co/meta-llama/Meta-Llama-3-70B-Instruct}}(Llama3-8B, Qwen, and Llama3-70B for short).
We choose medium-sized, open-source LLMs to reflect realistic deployment scenarios where large proprietary models may be too expensive or inaccessible.
Notably, the challenges addressed in this work stem from the gap between the generalized capabilities acquired through pretraining and the specialized competencies required for the table extraction task, which cannot be overcome solely by switching to a larger or more capable LLM.
The deep embedding model $Emb(\cdot)$ is SFR-Embedding-Mistral\footnote{\url{https://huggingface.co/Salesforce/SFR-Embedding-Mistral}}~\cite{meng2024sfrembedding}, a state-of-the-art open-source embedder.
The surrogate model $M(\cdot)$ is BART-Large\footnote{\url{https://huggingface.co/facebook/bart-large}}~\cite{lewis2020bart}.
For the Schema Prefix that ends with an enumeration of known attributes, we randomly sample 10 different permutations, use them to calculate $scs(\cdot)$, and take the maximum score for each candidate. We will elaborate on the robustness of this implementation in Section~\ref{sec:ar-exp}.

\revision{During validation, 100 samples are held out for early stopping and method tuning, and the remaining samples are used for training and retrieval. At inference, the validation set is also added for retrieval.}
The final number of shots is $k^{te}$=$k^{ar}$=$5$ for Incidents and Weather, \revision{$3$ for Conversation}; the final LLM decoding uses temperature $T^{llm}$=5 and top-$p^{llm}$=0.5.
All the experiments are conducted on a server equipped with Intel(R) Xeon(R) Gold 6240 CPU and one NVIDIA A800 (80GB Memory) for Llama3-8B and Qwen,  \revision{two NVIDIA A800s for Llama-70B.} 
Each experiment is repeated 3 times with different random seeds, and we report the average and standard deviation. Our codebase is available at \url{https://github.com/LITONG99/tear}.

\subsection{Q1: Table Extraction Results}
\label{sec:te-exp}
The TE performance is in Table~\ref{tab:te} with the following key observations.
(1) 
\modelname~is consistently the best. This demonstrates that our method effectively selects more informative demonstrations by leveraging both textual and tabular characteristics. 
Among the baselines, RAG dynamically retrieves demonstrations, which is better than MapMake that uses fixed demonstrations, showing the critical role of adaptive demonstrations.
%
(2) ICL paradigm methods exhibit a clear advantage over the supervised method, especially on structured‑F1 that evaluates values. This agrees with the intuition that while patterns in header regions are relatively limited and are easier to learn via supervised training, the values in n-texts vary considerably with the input text, requiring a stronger generalization capability for correct extraction. 
Furthermore, larger LLMs are overall better, and this gap is more pronounced for MapMake and RAG. This indicates that providing well‑selected examples with \modelname~can particularly boost the performance of relatively weaker LLMs.
\revision{(3) The results also indicate that our new benchmarks are more challenging for the studied extraction systems than Conversation, as it is reported to have generally lower results, and the performance gap between different methods is larger. 
This aligns with our claim that the variability of n-texts is a more urgent challenge for existing systems, instead of the verbosity or value drifts characterized by Conversation.
}

\begin{figure*}[htbp]
    \centering
    \includegraphics[trim=0 10 0 2, clip, width=\linewidth]{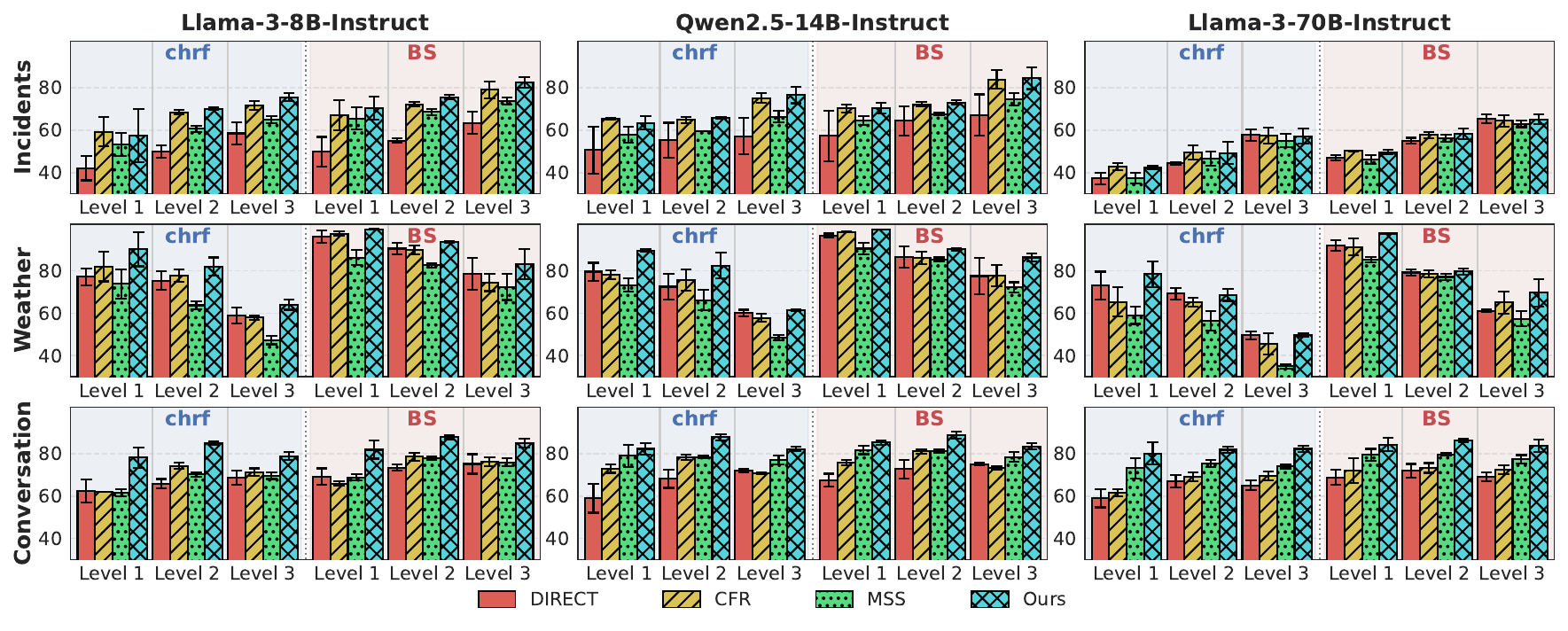}
    \caption{\revision{Attribute recommendation recall-AUC (\%).}}
    \label{fig:ar}
\end{figure*}
\begin{figure*}[htbp]
    \centering
    \includegraphics[trim=18 5 15 0, clip, width=\linewidth]{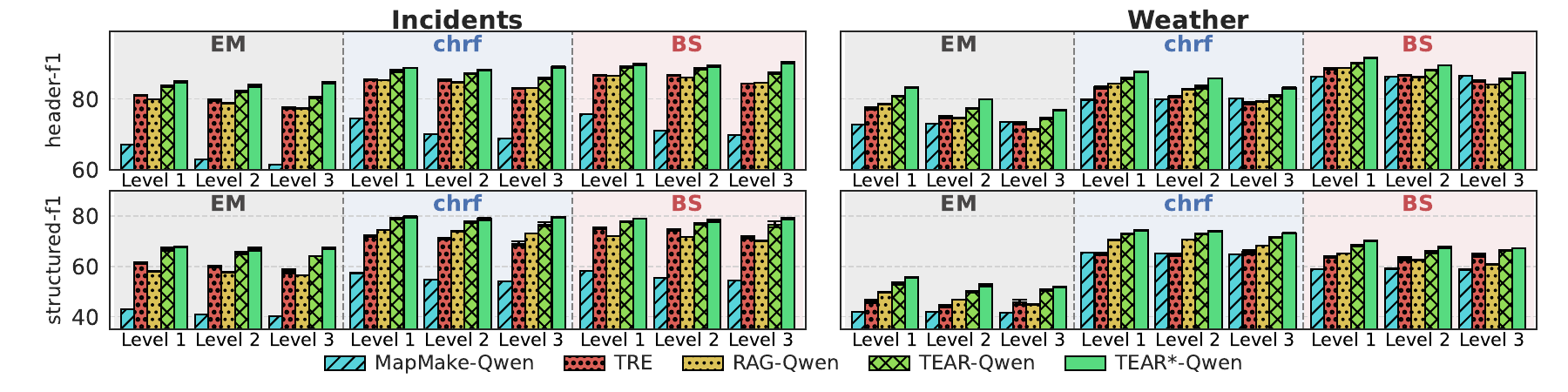}
    \caption{Table extraction with text-driven attributes with Qwen backbone.}
    \label{fig:tear-qwen}
\end{figure*}

\subsection{Q2: Attribute Recommendation Results}
\label{sec:ar-exp}
The attribute recommendation performance is shown in Fig.~\ref{fig:ar}. 
(1) The results confirm the viability of using LLMs for AR. Particularly, on the Weather dataset, where the recall-AUC reaches an exceptionally high level in some cases, indicating that the LLM-recommended attributes semantically encompass nearly all ground-truth attributes.
(2) Our method achieves the best overall performance. 
\revision{Under all comparisons, \modelname~is the best on 48 out of 54 comparisons. }
\revision{(3) After our method, the baselines do not have an obvious second best, and the LLM inference, global frequency, and semantic similarity have their own advanced cases.}
(4) Trends across exploratory levels differ for different datasets. Recall that Levels 1 to 3 aim to discover the remaining 20\%, 35\%, and 50\% of attributes.
On Weather, performance improves from Level 3 to Level 1, indicating that a more complete initial schema makes the task easier.
In contrast, Level 1 is the hardest for Incidents. Examining the schema splits, we find that this is due to a few low‑frequency attributes (e.g., “Number of rounds fired”) that are particularly difficult to discover precisely; consequently, the task becomes increasingly harder as the schema grows more complete.

\subsection{Q3: Table Extraction with Text-Driven Attributes}
\label{sec:exp-tear}
We simulate the process of users expanding the schema with recommended attributes for table extraction, in order to demonstrate the effect of exploratory schema design on the overall table extraction system. 
Specifically, we take $Z'[:k]\cap \hat{S'}$ as the user-accepted text‑driven attributes (that "checked" in Fig.~\ref{fig:source-text}(c)).
We numerically estimate $k$ by the elbow point~\cite{satopaa2011finding} of the $scs(\cdot)$, which reflects a realistic scenario where users only navigate top-ranked attributes. 
Moreover, using set intersection to select attributes (i.e., only adopting a recommendation if its name exactly matches the ground truth) is a very conservative simulation. In practice, users would typically accept a recommendation as long as it is semantically close to their interests and appropriately formulated. 
Under each exploratory scenario, we first evaluate table extraction performance by running TEW. We then execute the ARW, update, and run TEW again to obtain the updated performance, denoted as TEAR*. Both results are compared against the ground‑truth table following the complete schema. Results using Qwen are in Fig.~\ref{fig:tear-qwen}; those with other LLMs exhibit similar trends.
It shows that TEAR consistently outperforms the baselines, and the interactive pipeline TEAR* achieves additional performance gains. 
Notably, on the Incidents dataset, while other methods degrade as the exploratory level increases, \modelname* maintains stable performance and even shows improvement in some settings, highlighting the benefits of text‑driven attributes.

\subsection{Q4: Efficiency}
\begin{figure*}[ht]
    \centering
    \includegraphics[trim=110 0 80 0, clip, width=\linewidth]{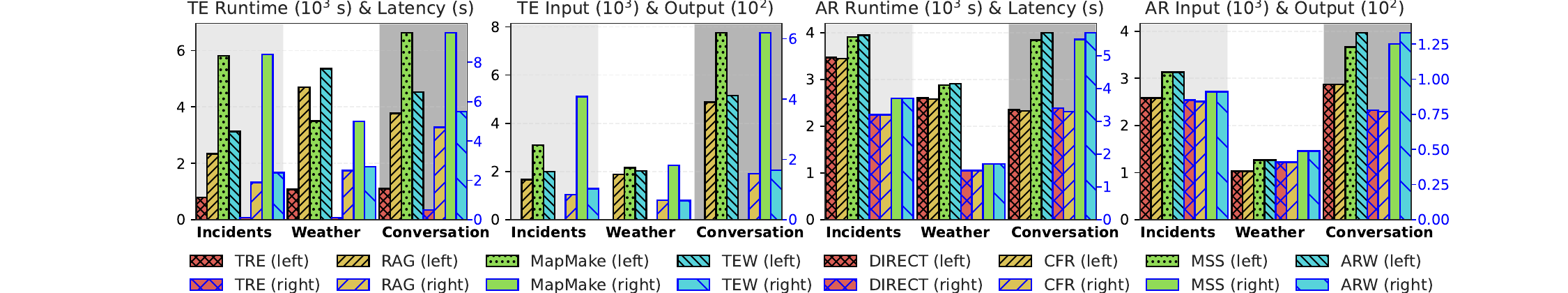}
    \caption{\revision{Efficiency comparison with Llama3-8B. Patterns for other LLMs are similar.}}
    \label{fig:eff-llama8}
\end{figure*}
\revision{
Fig.~\ref{fig:eff-llama8} compares efficiency, where time is measured by running the LLM on a local research server, applying no acceleration, and the Input and Output tokens per text are counted only for LLM methods, 
For TE, the efficiency order is TRE>RAG>Ours>MapMake. 
For AR, results are averaged over three exploratory levels, and the efficiency order is CFR>MSS>DIRECT>Ours.
}
The overhead of LLM methods primarily arises from the internals of the backbone, not the methods themselves. Compared to the high cost of fine‑tuning or the extensive annotation effort required for supervised extraction models, our approach achieves strong performance with significantly lower demand for labels or computational resources.
\begin{table*}[t]
  \footnotesize
  \setlength{\tabcolsep}{3.5pt}   
  \centering
  \begin{minipage}[t]{0.54\textwidth}
    \centering
    \caption{Hybrid Integration Strategy efficiency.}
    \label{tab:eff}
    \begin{tabular}{c|cccc|ccc}
    \toprule
    Inci. & $|Z|$ & $|P^*|$ & l. & t.(ms) & \#Calls & \textsf{True} & \textsf{False} \\
    \midrule
    1 & 207$\pm$3 & 182$\pm$15 & 4 & 17$\pm$10 & 81$\pm$7 & 29$\pm$6 & 52$\pm$4 \\
    2 & 203$\pm$4 & 130$\pm$1 & 4 & 13$\pm$8 & 79$\pm$6 & 23$\pm$2 & 56$\pm$7 \\
    3 & 218$\pm$13 & 157$\pm$20 & 4 & 22$\pm$11 & 91$\pm$6 & 23$\pm$2 & 68$\pm$6 \\
    \cline{1-8}
    \toprule
    Wea. & $|Z|$ & $|P^*|$ & l. & t.(ms) & \#Calls & \textsf{True} & \textsf{False} \\
    \midrule
    1 & 279$\pm$9 & 10$\pm$3 & 3 & 12$\pm$0 & 9$\pm$2 & 4$\pm$1 & 5$\pm$1 \\
    2 & 238$\pm$4 & 380$\pm$124 & 4 & 37$\pm$3 & 198$\pm$51 & 30$\pm$3 & 167$\pm$53 \\
    3 & 212$\pm$21 & 315$\pm$122 & 4 & 28$\pm$5 & 163$\pm$40 & 33$\pm$5 & 130$\pm$36 \\
    \cline{1-8}
    \toprule
    \revision{Con.} & \revision{$|Z|$} & \revision{$|P^*|$} & \revision{l.} & \revision{t.(ms)} & \revision{\#Calls} & \revision{\textsf{True}} & \revision{\textsf{False}} \\
    \midrule
    \revision{1} & \revision{94$\pm$4} & \revision{156$\pm$31} & \revision{6} & \revision{15$\pm$2} & \revision{10$\pm$0} & \revision{6$\pm$2} & \revision{4$\pm$1}\\
    \revision{2} & \revision{128$\pm$2} & \revision{173$\pm$7} & \revision{4} & \revision{26$\pm$4} & \revision{13$\pm$2} & \revision{8$\pm$1} & \revision{5$\pm$ 1} \\
    \revision{3} & \revision{134$\pm$7} & \revision{235$\pm$79} & \revision{7} & \revision{24$\pm$3} & \revision{25$\pm$8} & \revision{11$\pm$5} & \revision{14$\pm$6} \\
    \bottomrule
    \end{tabular}
  \end{minipage}
  \hfill
  \begin{minipage}[t]{0.45\textwidth}
    \centering
    \caption{Influence of Hybrid Integration Strategy with Qwen. 1, 2, and 3 are exploratory levels.}
    \label{tab:tear-i-qwen}
    \begin{tabular}{c|ccc}
    \toprule
    Dataset & $\Delta$ chrf & $\Delta$ BS & Ratio (\%) \\
    \midrule
    Inci. 1 & 0.009$\pm$0.004 & 0.005$\pm$0.011 & 16.4$\pm$1.9 \\
    Inci. 2 & 0.007$\pm$0.001 & 0.002$\pm$0.007 & 12.9$\pm$0.5 \\
    Inci. 3 & 0.005$\pm$0.006 & 0.003$\pm$0.006 & 12.4$\pm$1.3 \\
    \cline{1-4}
    Wea. 1 & 0.001$\pm$0.001 & 0.001$\pm$0.000 & 1.3$\pm$0.1 \\
    Wea. 2 & 0.002$\pm$0.015 & -0.007$\pm$0.009 & 13.9$\pm$0.6 \\
    Wea. 3 & -0.001$\pm$0.002 & 0.005$\pm$0.003 & 17.1$\pm$0.1 \\
    \cline{1-4}
    \revision{Con. 1} & \revision{0.002$\pm$0.001} & \revision{0.002$\pm$0.001} & \revision{8.4$\pm$3.3} \\
    \revision{Con. 2} & \revision{0.001$\pm$0.000} & \revision{0.001$\pm$0.000} & \revision{8.3$\pm$2.7} \\
    \revision{Con. 3} & \revision{0.001$\pm$0.000} & \revision{0.001$\pm$0.000} & \revision{12.0$\pm$5.2} \\
    \bottomrule
    \end{tabular}
  \end{minipage}
\end{table*}
We further evaluate the efficiency of the Hybrid Integration Strategy in Table~\ref{tab:eff}, where 1,2,3 are exploratory levels. Here, $|Z|$ is \#attribute proposals, $|P^*|$ is \#low-diversity sets, \textit{l.} is the size of the largest low-diversity set affecting pruning search iterations, \textit{t.} is the pruning time.
$|P^*|$ is efficiently small, showing that many proposals are actually far from others and can be quickly excluded from duplicate analysis. 
And thanks to our breadth-first greedy scheduling (Algorithm~\ref{alg:bfs}), the number of LLM calls is even less. 

\begin{figure*}[t]
  \centering
  \begin{minipage}[t]{0.49\textwidth}
    \centering
    \includegraphics[trim=0 10 0 2, clip, width=\linewidth]{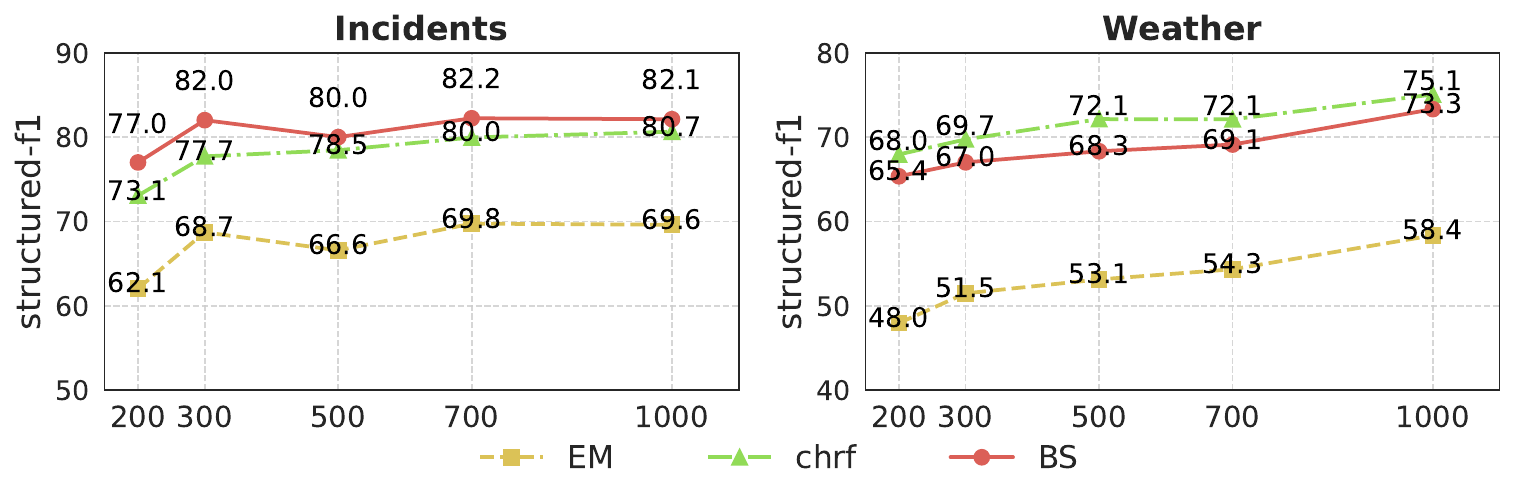}
    \captionof{figure}{\revision{Labeled pool size ablation with Llama3-8B.}}
    \label{fig:pool_size_ablation_llama}
  \end{minipage}
  \hfill
  \begin{minipage}[t]{0.49\textwidth}
    \centering
    \includegraphics[trim=0 10 0 2, clip, width=\linewidth]{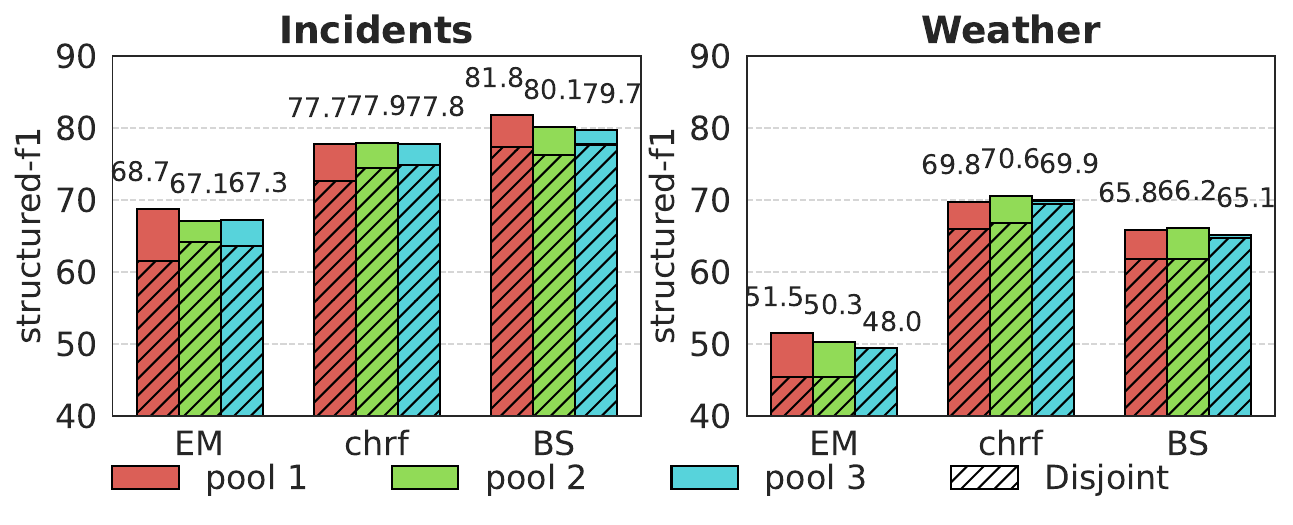}
    \captionof{figure}{\revision{Different labeled pools with Llama3-8B.}}
    \label{fig:disjoint-llama}
  \end{minipage}
\end{figure*}

\subsection{\revision{Q5: Ablation Studies}}
\subsubsection{\revision{Labeled Pool Size, Split and Initialization}}
\label{sec:ablation-pool}
We vary the size, split, and initialization of the labeled pool and report the structured-F1 with Llama3-8B in Fig.~\ref{fig:pool_size_ablation_llama} and Fig.~\ref{fig:disjoint-llama}, while other LLMs and metrics have the same pattern.
In Fig.~\ref{fig:pool_size_ablation_llama}, we reserve a subset of 200 test samples and progressively enlarge the labeled pools. 
The upward trend in Incidents gradually saturates, while Weather exhibits continued improvement, indicating a higher demand for labels.
In Fig.~\ref{fig:disjoint-llama}, we use the same reserved test set and sample three different pools, which lead to similar results, demonstrating the robustness of \modelname~ against initialization.
We further compare our default setting, where the same pool is used for learning $M$ and retrieval, with its \textbf{disjoint} setting, where the pool is split for learning and retrieval. 
The disjoint setting consistently underperforms our default setting, verifying that allocating a standalone retrievable set is unnecessary under data scarcity.

\subsubsection{\revision{Proactive Demonstration Module}}
\begin{figure*}[t]
  \centering
  \begin{minipage}[t]{0.49\textwidth}
    \centering
    \includegraphics[trim=0 10 0 2, clip, width=\linewidth]{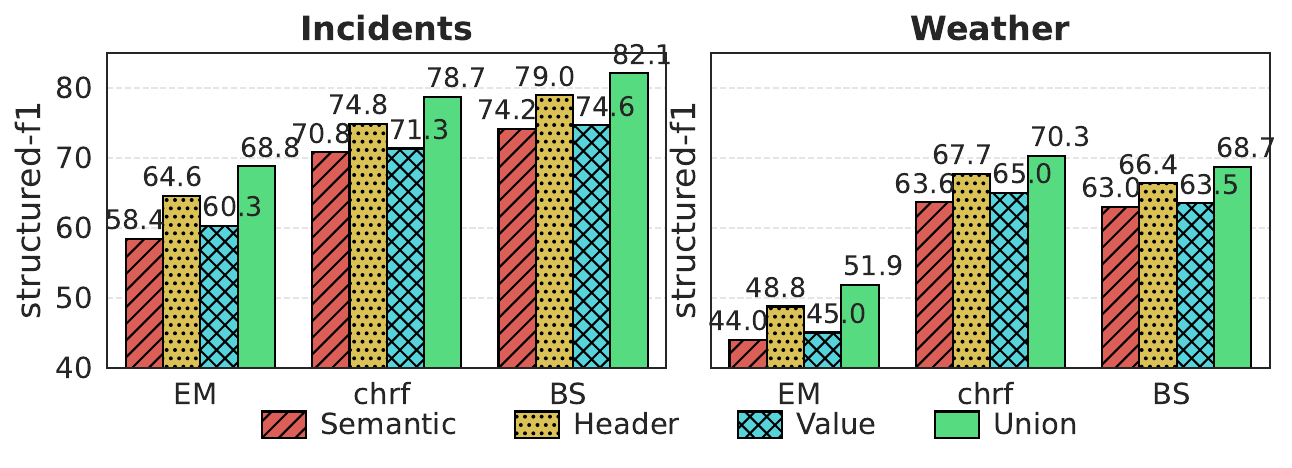}
    \captionof{figure}{\revision{Utility scores ablation with Llama3-8B.}}
    \label{fig:retrieve_ablation_llama}
  \end{minipage}
  \hfill
  \begin{minipage}[t]{0.5\textwidth}
    \centering
    \includegraphics[trim=0 10 0 2, clip, width=\linewidth]{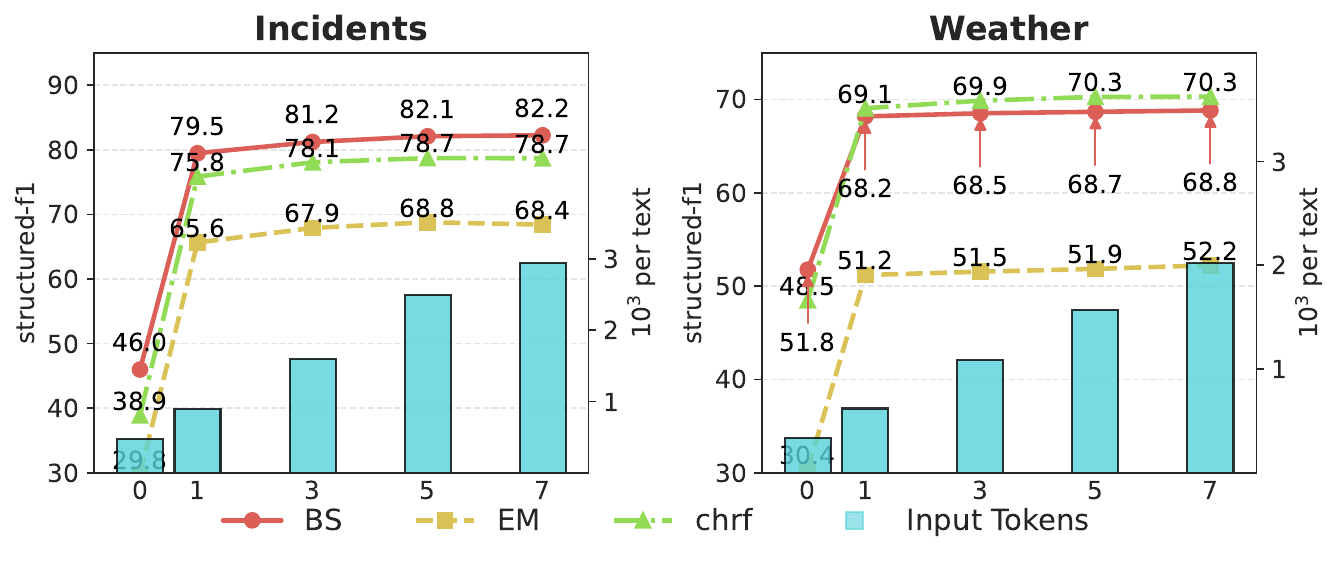}
    \captionof{figure}{\revision{Retrieve shots ablation with Llama3-8B.}}
    \label{fig:shot_ablation_llama}
  \end{minipage}
\end{figure*}

We ablate the demonstrations retrieved by each utility score, and the extraction performance is in Fig.~\ref{fig:retrieve_ablation_llama}, showing that the header and value utilities derived by previews are more beneficial than the plain text semantic, and the union of three utilities is the best. 
We also ablate the number of retrieved examples, $k^{te}$, in Fig.~\ref{fig:shot_ablation_llama}. Results show that only 1 shot could greatly boost the performance, and the performance reaches a high level after a few shots. 
The results of other backbones show similar patterns.

\begin{figure*}[th]
    \centering
    \includegraphics[trim=0 10 0 0, clip, width=\linewidth]{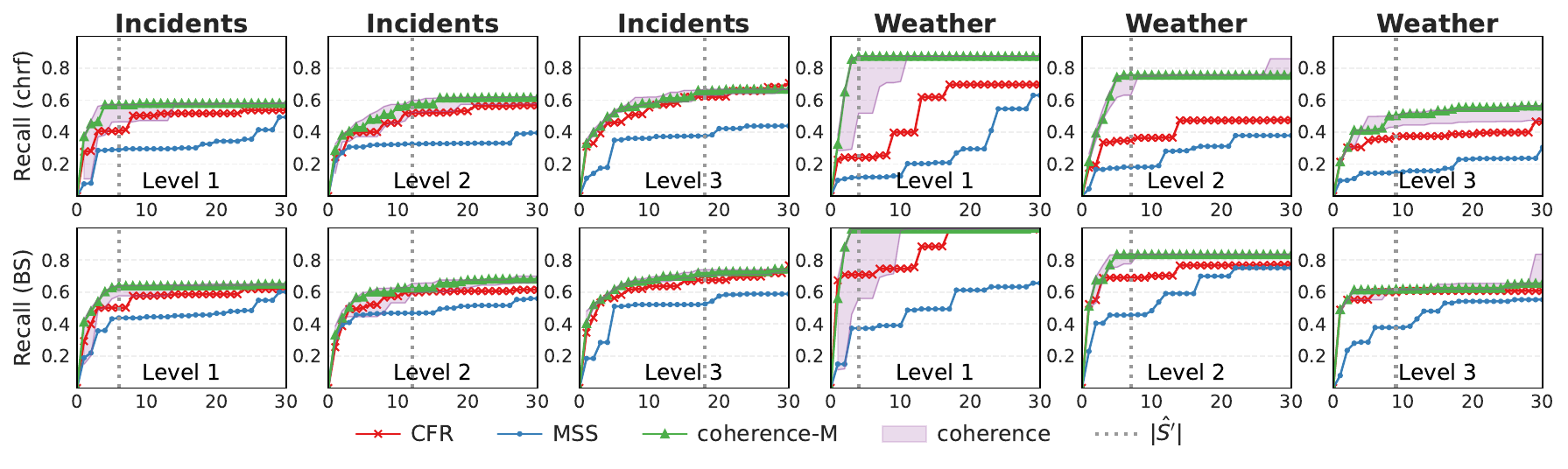}
    \caption{Recommendation recall of different ranking methods with Qwen.}
    \label{fig:ar-qwen}
\end{figure*}
\subsubsection{\revision{Hybrid Integration Strategy}}
\label{sec:ablation-his}
\begin{figure}[th]
    \centering
    \includegraphics[trim=0 10 0 2, clip, width=0.6\linewidth]{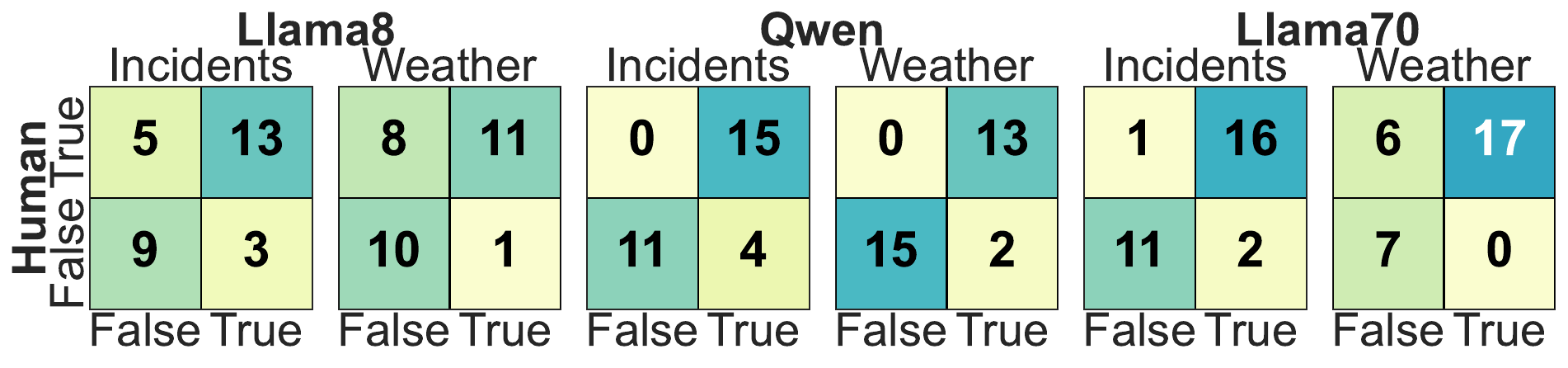}
    \caption{\revision{Agreements between LLMs and humans.}}
    \label{fig:manual-dedup}
\end{figure}
We report the end-to-end improvement of Hybrid Integration Strategy and the Ratio of removed proposals. 
Results with Qwen are in Table~\ref{tab:tear-i-qwen}, and those of other LLMs are similar. It shows that Qwen removes an average of 11\% proposals while maintaining comparable recall-AUC, and that ratio for Llama3-8B is 15\%, and Llama3-70B 7\%. 
Moreover, we find that the duplicates predominantly occur among low‑coherence proposals, which are at the tail of the recommendation list. 
This explains why integration brings only modest improvement on recall-AUC ($\Delta$s).
In order to evaluate the alignment between the LLM’s judgment on duplicates with human’s, we we sample 30 cases and expose the same context to a human for each setting, and results are shown in Fig.~\ref{fig:manual-dedup}.
Based on Cohen's Kappa~\cite{Landis1977TheMO}, Llama3-8B ($\kappa$=0.43-0.46) and Llama3-70 ($\kappa$=0.57-0.79) achieve moderate agreement, and Qwen ($\kappa$=0.73-0.87) achieves substantial agreement. It is hence a pragmatic design to appoint an LLM as the agent for proposal deduplication.

\subsubsection{Semantic Coherence Score}

To illustrate the quality of the highest‑ranked attributes, namely when the budget $k$ is small, we further plot the recall curves for the top‑k recommendations with Qwen in Fig.~\ref{fig:ar-qwen}, and those with other LLMs show similar patterns.
In contrast to the baselines, the results show that our Schema Coherence Score consistently pushes high‑quality candidates toward the top of the list, substantially improving early recall, which is ideal for practical short‑list applications. This difference is especially evident on the Weather dataset under Levels 1 and 2.

We further study the effect of varying the schema attribute permutations succeeding the SP, shown as the shaded region (e.g., there will be 6 possible permutations for 3 attributes).
This sensitivity arises because LLMs tend to attend more strongly to nearby context, the last few attributes, during next‑token prediction. 
To mitigate this sensitivity, our full implementation (coherence‑M) samples 10 random permutations and retains the maximum score for each candidate, as detailed in the implementation. This strategy successfully avoids underperforming orders and captures each candidate’s peak coherence across multiple contextualizations.
\subsubsection{\revision{Manual Evaluation on Structured-F1}}
\revision{We conduct a human evaluation on 200 predictions across all benchmarks. Human evaluators are asked to compare the original prediction with a minimally perturbed version introducing a single semantic difference: either single-value corruption or value swap of two cells. The direction of structured-F1 change aligns closely with human preference (accuracy 90\%, 95\%, 92.5\% for sim.=EM, chrf, BS.).}



\section{Conclusion}
\label{sec:conclusion}
We introduce \modelname, a framework that leverages LLM in‑context learning for table extraction with attribute recommendation. 
In the Table Extraction Workflow, we propose a Proactive Demonstration Module to customize demonstrative examples for each input, addressing the ineffectiveness of heuristic instructions by dynamically adapting to the high variability of n‑texts.
In the Attribute Recommendation Workflow, we design a Discovery Mechanism, a Hybrid Integration Strategy, and a Schema Coherence Score to openly discover, consolidate, and present a ranked list of text‑driven new attributes to the user, overcoming the limitations of fixed heuristic schemas. 
This recommendation capability is the most distinctive feature of our method, tackling the exploratory schema design challenge of naturally occurring texts. 
We further establish the evaluation benchmarks on n‑texts, including two new datasets, appropriate metrics, and a discussion of baselines. 
Experiments with \revision{three} open‑source LLMs show that \modelname~achieves state‑of‑the‑art performance on both tasks, and the recommended attributes effectively enhance table extraction in exploratory scenarios.

\section{Discussion, Limitation and Future Work}
\textbf{1.~AR validation and feedback loop.}
Our exploratory scenarios are created by removing low-frequency attributes from a human-designed complete schema, and we simulate the recommendation acceptance using such hidden ground truth. Therefore, the current evaluation shows the ability to recover omitted human-designed attributes under a controlled setting, but does not directly measure real user behavior. Evaluation with naturally incomplete schemas or real users would be important future directions.
Moreover, we clarify that AR does not have an inherent convergence point that iterative refinement could reach because the open‑ended candidate space is exhaustive if continuously prompted. As the first to establish the task, we focus on single-round quality, which provides known attributes to the system at the beginning.
Multi‑round interaction would require modeling the relationships among attributes provided across rounds and a more complex benchmark, which could be pursued in future work. 
\textbf{2.~LLMs as core components and other design choices.} 
Leveraging LLMs offers advantages for our tasks and is a common, well-established choice in recent works. Our contribution then lies in closing the gap between an LLM's general capability and the task-specific competence, which is a gap that a stronger general LLM or better prompts alone cannot close.
Our design choices are based on standard machine learning, established prior works, or empirical data statistics. Our contribution is not the scripted LLM instructions, but the data‑driven components and the framework.

\section*{Acknowledgments}
The research was supported in part by an NSFC grant 62432008, RGC RIF grant R6021-20, an RGC TRS grant T43-513/23N-2, RGC CRF grants C7004-22G, C1029-22G and C6015-23G, NSFC/RGC grant CRS\_HKUST601/24 and RGC GRF grants 16207423, 16203824 and 16216725.
Lei Chen's work is partially supported by National Key Research and Development Program of China Grant No.2023VFF0725100, NSFC under Grant No. U22B2060, Guangdong-Hong Kong Technology innovation Joint funding Scheme Project No. 2024A0505040012, the Hong Kong RGC GRF Project 16213620, RIF Project R6020-19, AOE Project AoE/E-603/18, Theme-based project TRS T41-603/20R, CRF Project C2004-21G, Key Areas Special Project of Guangdong Provincial Universities 20242DZX1006, Guangdong Province Science and Technology Plan Project 2023A0505030011, Guangzhou municipality big data inteligence key lab, 2023A03J0012, Hong Kong ITC ITF grants MHX/078/21 and PRP/004/22FX, Hong Kong ITC TC-SKLCRCC26EG01, Zhujiang scholar program 2021JC02X170, Microsoft Research Asia Collaborative Research Grant, HKUST-Webank joint research lab, 2025 HKUST Shenzhen-Hong Kong Collaborative Innovation Institute Green Sustainability Special Fund from Shui On Xintiandi and the InnoSpace GBA and HKUST(GZ)- CMCC(Guangzhou Branch) Metaverse Joint Innovation Lab under Grant No. P00659. 
Jiachuan WANG's work is supported in part by JST CREST (JPMJCR22M2).


\section{Appendices}
\subsection{Supplementary Discussion for Hybrid Integration Strategy}
\subsubsection{Vendi Score}
\label{app:vs}
Vendi Score is the effective rank of the normalized similarity matrix,
\begin{equation}
    vds(q) = \exp(-\text{tr}(\frac{\mathcal{K}_{q\times q}}{|q|}\log \frac{\mathcal{K}_{q\times q}}{|q|})),
\end{equation}where $\mathcal{K}_{q\times q}$ is the similarity matrix of size $|q|\times|q|$ and tr$(\cdot)$ means trace.
Vendi Score is maximized as $|q|$ if $\mathcal{K}(i,j)=0,\forall i,j\in q,  i\neq j$, meaning every two elements are totally different, and it is minimized as $1$ if $\mathcal{K}_{q\times q}$ contains all $1$s. 
Since $q\subseteq Z\cup S=B$. We precompute and cache $\mathcal{K}_{B\times B}$ to acquire any submatrix $\mathcal{K}_{q\times q}$ from it. Given $\mathcal{K}_{q\times q}$, the computational complexity is in $O(|q|^3)$~\cite{dan2023vendi}. We set $\mathcal{K}(a,b)=($chrf$\beta(a,b)$+chrf$\beta(b,a))/2$, reflecting pattern similarity as in Equation~\ref{value-sim}.

\begin{algorithm}[h]
\caption{Pruned Bottom-Up Search for $P^*$ }\label{alg:pruned}
\begin{algorithmic}[1]
\Require attribute proposals $Z$, known attributes $S$, threshold $\delta$
\Ensure $P^*$
\State $P^*=\emptyset$, Br$\leftarrow \{\}$, len$\leftarrow 1$ \Comment{Initialize dictionary.}
\For{$p\in Z\cup S$} Br$[p]=Z\cup S\setminus\{p\}$ \Comment{Start by len=1.} \EndFor 
\While{|Br|$>0\land$len$<|Z|+1$} \Comment{Outer loop.}
    \State nextBr$\leftarrow \{\} $
    \For{$p \in$ Br, $z' \in \text{Br}[e]$} \Comment{Inner loop.}
        \State valid $\leftarrow$ \textit{True}, activeBr $\leftarrow$ Br[$p$] 
        \For{$z\in p$}: \Comment{Check all subset of length $|p|$}
            \State $p'= p\setminus\{z\}\cup\{z'\}$ \Comment{Replace one element, $|p'|=|e|$}
            \If{$p'\in$ Br} activeBr $\leftarrow$ activeBr $\cap$ Br[$p'$] \Comment{Prune.}
            \Else \quad valid$\leftarrow False,$ \textbf{Break} 
            \EndIf
        \EndFor
        \If{valid $=$ \textit{True}}
            \Comment{All proper subset of $p$ is in $P^*$.}
            \State $p'=p\cup \{z'\}$
            \If{$vds(p')< 1+\delta$}
                $P^*\leftarrow P^*\cup\{p'\}$, nextBr$[p']\leftarrow$ activeBr  
            \EndIf
        \EndIf
    \EndFor
    \State len$\leftarrow$len$+1$, Br$\leftarrow$ nextBr \Comment{Search for the larger size}
\EndWhile
\end{algorithmic}
\end{algorithm}

\label{app:p-star}
\subsubsection{Pruned bottom-up search for low-diversity set}
As established by Equation~\ref{eq:div}, if a set is in $P^*$, then all of its subsets must also be in $P^*$. 
In other words, the monotonicity of $div(\cdot)$ means $div(p\cup\{z\})\geq div(p), \forall z\notin p$. 
Therefore, we compute $P^*$ in ascending order of size as in Algorithm~\ref{alg:pruned}. 
Specifically, in each iteration of the outer loop, we examine all qualified sets of size $\text{len}+1$. 
Dictionary $\text{Br}$ maps each qualified set of size len to a set of candidate attributes $\text{Br}[p]\subseteq Z\cup S$ that may be added to $p$ to form a larger qualified set. 
Within the inner loop, we consider the candidate extension $p' = p\cup\{z'\}$ for some $z'\in\text{Br}[p]$. 
It first performs a fast lookup to verify that replacing any element $z\in p$ with $z'$ yields a known qualified set. 
If all proper subsets of $p'$ are already known to be qualified,  $vds(p')$ is computed to decide the final answer.

\subsubsection{Complexity}
\label{app:bfs}
\underline{\emph{Diversity checking}}. Let $N = |Z \cup S|$ be the total number of candidate attributes, and $L = \max_{p \in P^*} |p|$ be the size of the largest low‑diversity set.
For $i \in \{1,\dots,L\}$, define $M_i = |\{p \in P^* : |p| = i\}|$, with $M_1 = N$.
Algorithm~\ref{alg:pruned} constructs $P^*$ by attempting to extend each qualified set of size $i$ with at most $N-i$ candidate attributes.
The total number of examined sets is therefore bounded by
$|P'| \leq \sum_{i=1}^{L} (N - i) M_i = N \sum_{i=1}^{L} M_i - \sum_{i=1}^{L} i M_i = N |P^*| - \sum_{p \in P^*} |p|.$    
Therefore, $|P'|$ scales linearly with the output size $|P^*|$.
Each candidate extension incurs $O(i)$ dictionary lookups, and only a small fraction proceed to Vendi Score computation, yielding an overall time complexity of $O\bigl( \sum_{i=1}^{L} (N - i) M_i \cdot (i + C_{vds}) \bigr)$.
In practice, pruning in the inner loop further restricts the branch set $\text{Br}[p]$: the number of viable extensions for a set $p$ never exceeds that of any of its subsets, and Br$[p]$ shrinks as $p$ is extended to tighten the factor $(N-i)$. 
Empirically, most candidate attributes do not form duplicate pairs with any other proposal and only a small fraction participate in true redundancies. 
So, the factor is far smaller than $N$, and $L$ remains modest (Table~\ref{tab:eff}).
\underline{\emph{Contextualized analysis}}. Suppose there are $N'$ duplicates that can potentially be eliminated by a \textsf{True} response, and $|B| - N'$ non-duplicate elements. 
Then there must exist sets $q_1, q_2,\dots, q_{|B|-N'}$ in $P^*$ that together cover all elements, because every duplicate must co-occur with at least one non-duplicate in some set.
Consequently, for the set cover instance defined by the collection $P^*$ and the ground set $B$, the optimal solution size is strictly less than $|B| - N'$. A greedy set cover algorithm achieves an approximation ratio of $(\ln |B| + 1)$~\cite{chvatal1979greedy}, so the number of LLM calls incurred by $Q$ is at most $(\ln |B| + 1)(|B|-N')$. Line~10 further eliminates trivial low-diversity set.
Given the same $B$ and $|P^*|$, in the worst case, the algorithm traverses $P^*$ and receives \textsf{False} for all queries without performing any deduplication.
Such a situation is rare because $P^*$ is already of very low diversity.
The set cover analysis indicates that a greater number of duplicates allows an outer iteration to complete more quickly, which submits each element at least once.
We prioritize preserving the breadth of the search because the ultimate effect of deduplication hinges on whether each duplicate element is eliminated. 
The greedy heuristic actually processes larger sets earlier, helping the algorithm eliminate confirmed duplicates with few queries and shrinks the search space for subsequent iterations.

\subsubsection{Discussion of the ground truth.}
While resolving duplicate attribute proposals is essential, this step lacks ground truth, because semantic equivalence between attributes is inherently subjective. 
As for any recommendation system, the user's ultimate preference may deviate from what the history infers. 
Recall the earlier example of ``Medical center name'' versus ``Hospital name''. It is logical for a method to infer from the existing schema to merge them. Yet a user might insist on retaining both, much like a consumer abruptly deviating from past behavior to purchase an entirely different product. 
Given this inherent uncertainty, our method is pragmatic rather than optimized: we defer to the diversity metric and LLM's judgment to perform moderate processing for obvious duplicates, without pursuing a perfect formulation or solution. Our experiments further support this stance (Section~\ref{sec:ablation-his}).

\subsection{Data Annotation}
\label{app:data}
\begin{figure*}[th]
  \centering
  \begin{minipage}[t]{0.5\textwidth}
    \centering
    \includegraphics[trim=20 135 0 100, clip, width=\linewidth]{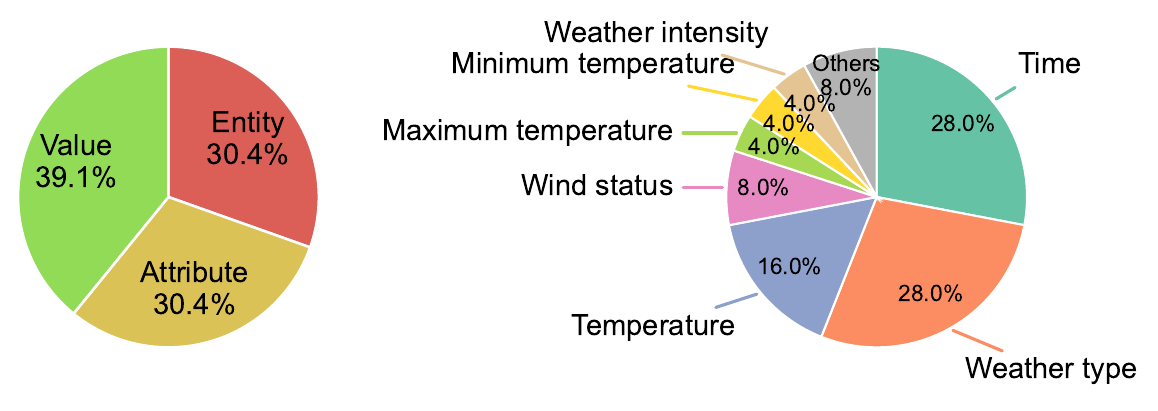}
    \captionof{figure}{Disagreement types on Weather.}
    \label{fig:disagreement_weather}
  \end{minipage}%
  \begin{minipage}[t]{0.5\textwidth}
    \centering
    \includegraphics[trim=20 135 0 100, clip, width=\linewidth]{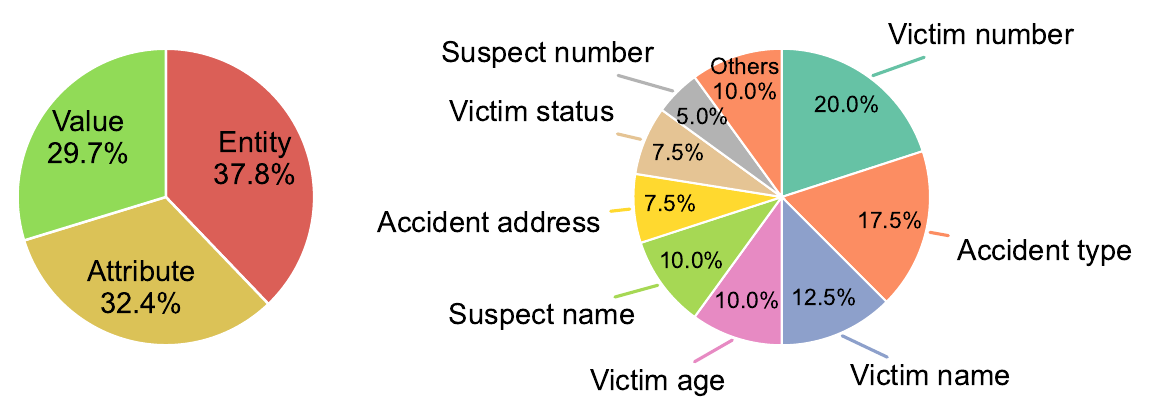}
    \captionof{figure}{Disagreement types on Incidents.}
    \label{fig:disagreement_incidents}
  \end{minipage}
\end{figure*}
We first deduplicate the original texts and remove corrupted texts. Three annotators (A, B, and C) participate in creating the table annotations, which proceeds in three steps for each dataset. First, the three annotators read the original CACAPO documentation and discussed the original attribute set to reach a consensus on the overall schema. 
Second, we sample 100 texts to be independently annotated by A, B, and C. 
Each annotator is provided with the original attribute‑value pairs and the complete table extraction schema. 
They are asked to fill in missing values, correct errors in the original annotations, and organize the correct values into multiple records to form a complete table. 
Disagreements are discussed until consensus is reached. 
The remaining texts are split into two parts and assigned to B and C, respectively. From each part, we sample 100 texts for A to annotate to evaluate agreement.  
The agreement between A and B is 88\%, and between A and C is 82\%. 
We categorize observed disagreements into three hierarchical levels, from coarse to fine-grained:
(1) \textbf{Entity-level}: Disagreements on the presence of entities, corresponding to how many rows a table should contain.
(2) \textbf{Attribute-level}: Disagreements on which attributes are instantiated for a given entity, without any Entity-level disagreement. This corresponds to determining the non-empty columns within a table row.
(3) \textbf{Value-level}: Disagreements on the value of a specific attribute, without any Entity- or Attribute-level disagreement. This corresponds to the cell content.
We analyze the distribution of disagreement types and the attribute-level breakdown of where disagreements occur, visualized in Fig.~\ref{fig:disagreement_weather} and \ref{fig:disagreement_incidents}.

\subsection{Variability Comparison Between Datasets}
\label{app:variability}
\begin{figure*}[thbp]
  \centering
  \begin{minipage}[t]{0.46\textwidth}
    \centering
    \includegraphics[trim=0 0 0 0, clip, width=\linewidth]{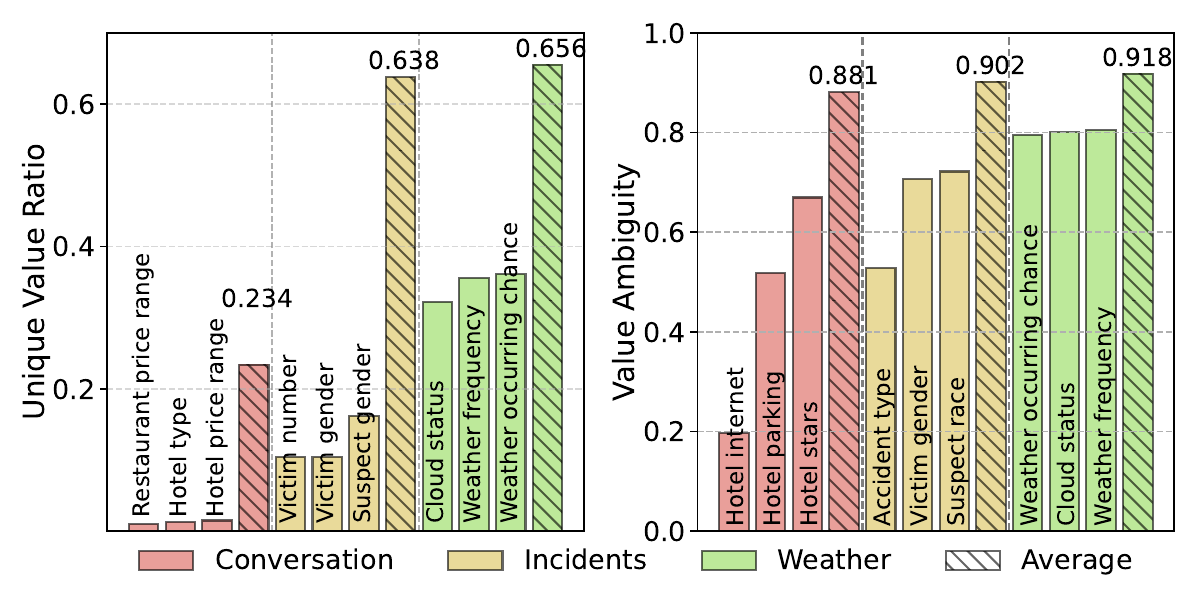}
    \caption{Attribute-level variability comparison.}
    \label{fig:variability}
  \end{minipage}
  \begin{minipage}[t]{0.54\textwidth}
       \centering
    \includegraphics[trim=0 0 5 0, clip, width=\linewidth]{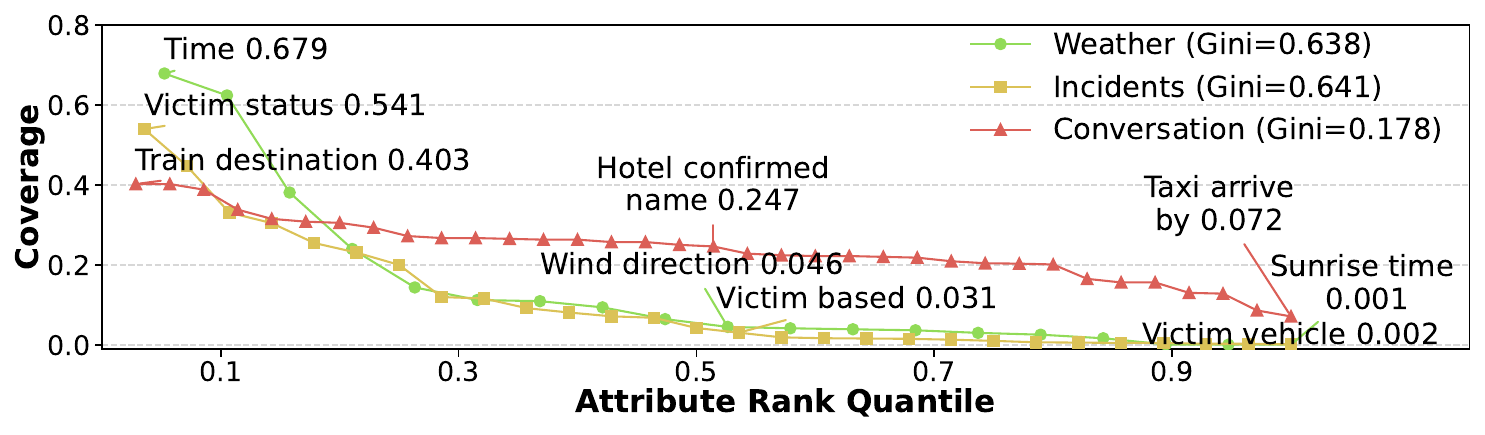}
    \caption{Attribute Coverage ranked curve.}
    \label{fig:coverage_curve}
  \end{minipage}
\end{figure*}

\revision{We conduct a statistical examination of the datasets' \textbf{variability} in two aspects: how an attribute is manifested in the text (Challenge 1) and which attributes are mentioned in a text (Challenge 2).}
\revision{To underscore Challenge 1, we compare the attribute-level variability:
 (1) Unique Value Ratio, $\in(0, 1]$: \#unique values / \#values. 
A higher score means the attribute manifests more unique values among all its mentions in the text. (2) Value Ambiguity $\in [0, 1]$: the normalized entropy of value distribution. A higher score indicates that no single value dominates the corpus, and the attribute lacks a global default interpretation. Fig.~~\ref{fig:variability} shows the average and the lowest scores (the ``easiest'') for each dataset. 
The former demonstrates the general difficulty while the latter reveals the lower-bound difficulty. These statistics imply the difficulty order of the datasets: Weather > Incidents > Conversation, aligning with the observation in Table~\ref{tab:te}. } For Challenge 2, we depict the Attribute Coverage (i.e., the ratio of texts mentioning the attribute) in Fig.~\ref{fig:coverage_curve}.
Apparently, the long-tail phenomenon of our new benchmarks is more significant with higher Gini than the Conversation dataset, reflecting the fact that Incidents and Weather contain more rare attributes.

\subsection{Computational Complexity of Structured-F1}
\label{app: f1}
Let $D$ has $n$ rows and $m$ columns while $\hat{D}$  has $\hat n$ and $\hat m$. Let $c$ denote the amortized cost to calculate the cell similarity.
To compute DF1 for some record pair, it requires $O(m\hat{m}c)$ to get the similarity matrix between header names, as in Fig.~\ref{fig:metric}(1), and $O((m+\hat{m})c)$ to average the similarity of matched values.
We compute DF1 for all $n\hat{n}$ record pairs with the $f_k$ similarity matrix reused, so the overall complexity for structured-F1 is $C1=O(m\hat{m}c+n\hat{n}(m+\hat{m})c)$. %
In contrast, if the table is flattened, there are $mn$ and $\hat{m}\hat{n}$ triples. So, the complexity of comparing triples is $C2=O(mn\hat{m}\hat{n}c)$.
For a reasonable system, the size of $D$ resembles that of $\hat D$, thus $m\approx\hat{m}$ and $n\approx\hat{n}$, so $C1=O(m^2c+2n^2mc)$, $C2=O(m^2n^2c)$.
When $n^2$>1 and $m$>3, $C1\leq C2$ and structured-F1 is more efficient; in other cases, $C1\approx C2$, so our metric does not increase the evaluation complexity.
\newpage
\bibliographystyle{ACM-Reference-Format}
\bibliography{tear}
\appendix
\end{document}